\pdfoutput=1
\documentclass[tikz,11pt,english,a4paper]{article}
\usepackage{float}
\usepackage{jcappub}
\usepackage[utf8]{inputenc}
\DeclareUnicodeCharacter{2212}{\ensuremath{-}}
\usepackage{subcaption}
\usepackage{mathtools}

\usepackage{bm}
\usepackage{nicefrac}
\usepackage{amssymb}
\usepackage{booktabs}
\usepackage{multirow}
\usepackage{placeins}
\usepackage{hyperref}

\newcommand{\dd}{\mathrm d}

\newcommand{\client}{CLiENT}
\newcommand{\clienturl}{\url{https://github.com/AarhusCosmology/client-f}}
\newcommand{\class}{\textsc{class}}

\newcommand{\cobaya}{Cobaya}
\newcommand{\PDP}{PDP+}
\newcommand{\HWC}{\ensuremath{H_0{\rm WC}}}
\usepackage{siunitx}

\usepackage[T1]{fontenc}
\usepackage{lmodern}
\rmfamily
\DeclareFontShape{T1}{lmr}{b}{sc}{<->ssub*cmr/bx/sc}{}
\DeclareFontShape{T1}{lmr}{bx}{sc}{<->ssub*cmr/bx/sc}{}

\newcommand{\CC}{C\nolinebreak\hspace{-.05em}\raisebox{.4ex}{\tiny\bf +}\nolinebreak\hspace{-.10em}\raisebox{.4ex}{\tiny\bf +}}
\newcommand{\CLASSpp}{\textsc{class}\nolinebreak\hspace{-.05em}\raisebox{.4ex}{\tiny\bf +}\nolinebreak\hspace{-.10em}\raisebox{.4ex}{\tiny\bf +}}

\makeatletter
\g@addto@macro\bfseries{\boldmath}
\makeatother

\newcommand{\e}{\mathrm{e}}
\newcommand{\Aosc}{A_\mathrm{osc}}
\newcommand{\kstar}{k_*}
\newcommand{\Mpl}{M_\mathrm{Pl}}
\newcommand{\nrun}{\alpha_\mathrm{s}}
\usepackage{needspace}

\begin{document}

\title{Features investigated: A detailed study of axion monodromy inflation using \client{}}

\author[a]{Steen~Hannestad,}
\author[a]{Rasmus~Lumbye,}
\author[b]{Andreas~Nygaard,}
\author[a]{Thomas~Tram}
\affiliation[a]{Department of Physics and Astronomy, Aarhus University,
 DK-8000 Aarhus C, Denmark}
\affiliation[b]{Department of Astrophysics, University of Zurich,
 CH-8057 Zurich, Switzerland}

\emailAdd{steen@phys.au.dk}
\emailAdd{andreas.hansen@uzh.ch}
\emailAdd{thomas.tram@phys.au.dk}

\abstract{
We investigate features in the primordial power spectrum from inflation, specifically those predicted by the axion monodromy model, using recent observational data.
In order to compute posteriors, profile likelihoods and Bayesian evidence accurately for the complex likelihood surfaces arising in this model we have used the \client{} emulator framework. As our data combinations we use both a ``conservative'' set consisting of Planck PR4 temperature, polarisation and lensing data together with DESI DR2 BAO and Pantheon+ supernovae, and a more ``aggressive'' set which also includes small-scale CMB data from ACT and SPT.
We find no robust oscillatory feature in current data. With the conservative data the best fit lies at a frequency $\omega \simeq 51$ and lowers $\chi^2$ by 10.1 relative to a featureless spectrum. However, this feature is disfavoured by the more aggressive data combination, whose best fit lies at $\omega \simeq 31$ and lowers $\chi^2$ by 7.4. Fits to featureless data are known to give similar improvements, and the Akaike information criterion, the Bayes factors and the look-elsewhere effect all indicate that neither improvement is significant. The amplitude is limited to $\Aosc < 0.056$ and $\Aosc < 0.062$ (95\%) for the two datasets. The emulators reproduce the true log-likelihood at samples of their posteriors with rms errors of 0.07 and 0.16, and the profile likelihoods are calculated with the true likelihood, using the profile capability of \client{}.}

\maketitle

\section{Introduction}
\label{sec:introduction}

Inflation predicts a primordial power spectrum of curvature perturbations that is very close to a power law, and the cosmic microwave background (CMB) measures it with percent precision over about three decades in wavenumber. Departures from a power law, so-called features, would carry information about the inflaton potential and about the physics from which it derives. Axion monodromy is a concrete example. It realises large-field inflation in string theory with an axion whose shift symmetry is broken by branes or fluxes, so that the field can traverse many periods of its potential \cite{Silverstein:2008sg,McAllister:2008hb,Kaloper:2008fb,Kaloper:2011jz}. Non-perturbative effects add a small periodic modulation to the potential. When the oscillation of the background is fast compared with the Hubble rate, the modulation drives a resonance in the mode functions, and the result is oscillations in the power spectrum that are periodic in $\ln k$ \cite{Flauger:2009ab,Flauger:2010ja}, together with a resonant bispectrum \cite{Chen:2008wn,Flauger:2010ja,Hannestad:2009yx,Chen:2010bka,Leblond:2010yq,Behbahani:2011it,Behbahani:2012be}.

These logarithmic oscillations have been searched for in every CMB data release since WMAP5, and they have not been detected. After a first fit of a modulated potential \cite{Pahud:2008ae}, fits of the monodromy template to WMAP data lowered $\chi^2$ by 11 to 20 at the cost of three additional parameters \cite{Flauger:2009ab,Aich:2011qv,Meerburg:2011gd,Peiris:2013opa}. However, fits to featureless simulated skies gave similar improvements \cite{Flauger:2009ab,Meerburg:2013cla}. The Planck analyses found best fits at $\omega \approx 30$ with temperature data and at $\omega \approx 18$ with polarisation added \cite{Planck:2013jfk,Planck:2015sxf,Planck:2018jri}, and calibrations against simulations put such improvements well within what noise produces: fits to noise typically gain $\Delta\chi^2 \approx 10$ \cite{Easther:2013kla,Meerburg:2013cla}. Searches for a drifting frequency \cite{Flauger:2014ana,Planck:2015sxf} and for the resonant bispectrum \cite{Planck:2013wtn,Planck:2015zfm,Planck:2019kim,Fergusson:2014hya,Munchmeyer:2014nqa,Munchmeyer:2014cca,Meerburg:2015owa} were null as well. Recent analyses combine Planck with ACT and SPT-3G \cite{Antony:2024vrx,Peng:2025vda,Nerval:2026iev}, reanalyse the Planck PR3 and PR4 spectra with all nuisance parameters free \cite{Raffaelli:2026bhk}, or add galaxy clustering from BOSS, eBOSS and DESI \cite{Zeng:2018ufm,Beutler:2019ojk,Ballardini:2022wzu,Mergulhao:2023ukp,Calderon:2025xod,Chen:2026gxp}. In galaxy clustering the oscillations are damped by the non-linear evolution of structure, and this damping has been modelled in perturbation theory \cite{Ballardini:2024dto} and calibrated with $N$-body simulations \cite{Cohen:2026pat}. These analyses limit the amplitude to roughly 2--3\% at 95\%~CL and find no significant signal once the look-elsewhere effect is taken into account \cite{Mergulhao:2023ukp,Raffaelli:2026bhk,Chen:2026gxp}. Free-form reconstructions of the primordial spectrum from Planck, ACT and SPT data find no significant deviation from a power law either \cite{Raffaelli:2025kew,Chandra:2026byw}. A search by symbolic regression selects instead an oscillation periodic in $1/k$, both in the Planck data and in its combination with ACT and SPT-3G \cite{Peng:2026kyq}. It fits the data better than the linear and logarithmic templates, with a weak preference for a non-zero amplitude. A search for the localised features that a burst of particle production during inflation imprints on the CMB finds a mild hint, at about $2\sigma$, in the combination of Planck and ACT \cite{Jense:2026haw}. In the CMB such primordial features are not significantly degenerate with signals of post-inflationary origin, such as early dark energy or an injection of energy into the primordial plasma, which do not generically reproduce the oscillations of sharp and resonant features \cite{Braglia:2026njc}.

Two difficulties are common to all these analyses. The first is numerical. Every evaluation of the likelihood requires a run of an Einstein--Boltzmann code with dense sampling of the primordial spectrum, and the likelihood has many narrow maxima in frequency and phase. Markov chains get trapped in them \cite{Flauger:2009ab,Flauger:2014ana}, standard minimisers perform poorly \cite{Planck:2015sxf}, and nested sampling under-resolves narrow maxima \cite{Hamann:2021eyw}. Most analyses have therefore fixed the $\Lambda$CDM or the nuisance parameters, or both, at least in part of the work. The second difficulty is statistical. The Bayesian evidence for an oscillation depends strongly on the prior ranges of amplitude and frequency \cite{Planck:2013jfk,Zeng:2018ufm,Hergt:2021qlh}, the marginalised limits depend on the frequency range \cite{Calderon:2025xod}, and a frequentist significance requires a look-elsewhere correction \cite{Gross:2010qma,Bayer:2020pva,Raffaelli:2026bhk}. Profile likelihoods are free of the prior volume effects of marginalisation \cite{Herold:2021ksg,Holm:2022kkd,Herold:2024enb}, and dedicated codes exist for them \cite{Holm:2023uwa,Nygaard:2023cus}. However, for logarithmic oscillations the published profiles are envelopes of nested-sampling chains or were computed with the other parameters fixed \cite{Planck:2018jri,Hamann:2021eyw}.

Accelerated methods have so far addressed only parts of the problem. Gaussian-process Bayesian optimisation needs about 1400 likelihood evaluations for the logarithmic template, but it fixes the $\Lambda$CDM and nuisance parameters and gives neither posteriors nor evidences \cite{Hamann:2021eyw}. A Gaussian-process emulator of the likelihood trained by Bayesian optimisation gives the Bayesian evidence with about a thousand times fewer likelihood evaluations than nested sampling, but it has so far been applied to $\Lambda$CDM and to its extension with spatial curvature \cite{Cohen:2026iij}. Normalising-flow samplers reduce the number of evaluations but still call the Einstein--Boltzmann code at every step \cite{Karamanis:2022alw,Nerval:2026iev}. Emulators of the CMB spectra \cite{Nygaard:2022wri,SpurioMancini:2021ppk} have not been applied to oscillatory features. Features that shift in $\ell$ as the parameters change are very hard to compress \cite{Jense:2024llt}, and a logarithmic oscillation does exactly that.

In this work we use the likelihood emulator \client{} \cite{Janken:2025wlq}, which learns the likelihood of the data directly as a function of all model and nuisance parameters, so that the emulator must learn the oscillation itself rather than its imprint on the spectra. \client{} is publicly available at \clienturl{}, including all the features used in this work. Since the version described in ref.~\cite{Janken:2025wlq}, the training has been extended so that it checks the maxima of the emulator against the true likelihood, decides when to stop from new evaluations of the true likelihood, and refines the emulator on samples of its own posterior. If this posterior has separate modes, i.e.\ regions of high probability between which a Markov chain does not move, the code finds them and gives them the same weight in the refinement. In each step of this refinement several networks are trained, and the one that agrees best with the true likelihood at samples of its own posterior is kept. Furthermore, emulators trained on different parts of the prior volume can be combined into a single one, and the code includes tools for the Bayesian evidence, which is computed with the nested sampler BEST \cite{Nygaard:2026fgl,Nygaard:2026hbr}, and for profile likelihoods refined with the true likelihood. This is the first application of these features, and we therefore describe them in some detail and state where they were used. We obtain the posterior distributions with all parameters free, compare them with the true likelihood, compute the Bayesian evidence, and calculate profile likelihoods of the oscillation parameters with the true likelihood. We use two combinations of current data: the full Planck temperature, polarisation and PR4 lensing data together with DESI DR2 BAO and Pantheon+ supernovae, which we call \PDP{}, and the data combination of the $H_0$ World Cup \cite{Schoneberg:2026eys,Schoneberg:2026vaf}, which we call \HWC{}, in which the small-scale CMB spectra come from ACT DR6 and SPT-3G. Since a slow oscillation resembles a change of the tilt and of the running of the spectrum, we allow the spectral index to run. We consider frequencies up to $\omega = 200$, which requires a much denser numerical sampling of the CMB spectra than the defaults of the Einstein--Boltzmann code, and we have checked this sampling against a converged one. Section~\ref{sec:theory} reviews the model and the template, Sec.~\ref{sec:numerics} describes the numerical setup, and Sec.~\ref{sec:results} presents the posteriors, the profile likelihoods and the model comparison, first for \PDP{} and then for \HWC{}, i.e.\ with the small-scale CMB data of ACT and SPT added. Section~\ref{sec:conclusions} concludes. Appendix~\ref{app:client} describes the steps of \client{}, the accuracy of the emulators and the calculation of the profiles in more detail, and Appendix~\ref{app:classpp} the precision of the Einstein--Boltzmann code.

\section{Model framework}
\label{sec:theory}

\subsection{Axion monodromy inflation}
\label{sec:theory:monodromy}

In axion monodromy the inflaton $\phi$ is an axion whose potential is not periodic: the coupling to branes or fluxes unwraps it into a monotonic function of $\phi$, of the form $V \propto \phi^p$ at large field values. Explicit constructions gave $p = 2/3$ \cite{Silverstein:2008sg} and $p = 1$ \cite{McAllister:2008hb}, and backreaction of the heavy fields flattens potentials to $p < 2$ in general \cite{Dong:2010in}. Instanton effects, and the leading corrections in four-form realisations \cite{Kaloper:2008fb,Kaloper:2011jz}, add a periodic term. For the linear potential \cite{Flauger:2009ab}
\begin{equation}
    V(\phi) = \mu^3 \left[ \phi + b f \cos\left(\frac{\phi}{f}\right) \right],
    \label{eq:potential}
\end{equation}
where $f$ is the axion decay constant and $b < 1$ keeps the potential monotonic. Here and below field values are in units of the reduced Planck mass $\Mpl$.

The smooth part of the potential determines the slow roll. For $V \propto \phi^p$ the number of e-folds between the time when the pivot scale $\kstar$ leaves the horizon and the end of inflation is $N_* \simeq \phi_*^2/(2p)$, and the spectral index and the tensor-to-scalar ratio are $n_\mathrm{s} - 1 \simeq -(p+2)/(2N_*)$ and $r \simeq 4p/N_*$. For $p = 1$ and $N_* = 55$ this gives $\phi_* \simeq 10.5$, $n_\mathrm{s} \simeq 0.973$ and $r \simeq 0.07$, which is excluded by the BICEP/Keck limit $r < 0.036$ at 95\%~CL \cite{BICEP:2021xfz}. We do not impose these relations between $n_\mathrm{s}$, $r$ and the potential. The smooth spectrum is a power law with a running of the spectral index, in which the amplitude, the spectral index and the running are free parameters, and tensor modes are not included. Our constraints on the oscillation therefore do not depend on the power $p$ of the smooth potential.

The periodic term makes the velocity of the inflaton oscillate with frequency $\dot\phi/f$. For the linear potential $\dot\phi/H \simeq -1/\phi$, so that the oscillation is fast compared with the expansion rate when $f\phi_* \ll 1$. Each mode is then driven resonantly while it is still inside the horizon, when its physical frequency matches the frequency of the background, and the resulting power spectrum is \cite{Flauger:2009ab,Flauger:2010ja}
\begin{equation}
    \Delta^2_\mathcal{R}(k) = \Delta^2_\mathcal{R}(\kstar) \left(\frac{k}{\kstar}\right)^{n_\mathrm{s}-1} \left[ 1 + \delta n_\mathrm{s} \cos\left(\frac{\phi_k}{f} + \Delta\varphi\right) \right],
    \qquad \phi_k \simeq \phi_* - \frac{\ln(k/\kstar)}{\phi_*},
    \label{eq:monodromy_spectrum}
\end{equation}
where $\phi_k$ is the field value at which the mode $k$ leaves the horizon and $\Delta\varphi$ is a phase that depends on the history of reheating and on microscopic details. The spectrum oscillates in $\ln k$ with the dimensionless frequency and amplitude
\begin{equation}
    \omega = \frac{1}{f\phi_*},
    \qquad
    \delta n_\mathrm{s} = 3b\left(2\pi f\phi_*\right)^{1/2} = 3b\sqrt{\frac{2\pi}{\omega}}.
    \label{eq:omega_f}
\end{equation}
The frequencies studied here, $1 \leq \omega \leq 200$, correspond to decay constants between about $5\times10^{-4}$ and 0.1, and the best fit with \PDP{} (Sec.~\ref{sec:results}) corresponds to $f \simeq \ensuremath{1.9\times10^{-3}}$ and $b \simeq 0.023$.

Two effects limit this description. Backreaction of the moduli makes $f$ drift along the trajectory, and with it the frequency \cite{Flauger:2014ana}. Neither Planck 2015 nor Planck 2018 found evidence for such a drift \cite{Planck:2015sxf,Planck:2018jri}, and we keep $\omega$ constant. At high frequency the effective field theory of the oscillating background becomes strongly coupled, for $\omega \gtrsim 200$ \cite{Easther:2013kla,Flauger:2014ana}, i.e.\ at the upper bound of the range considered here. At low frequency the resonance approximation $f\phi_* \ll 1$ no longer holds, and there the template is a phenomenological description. For $\omega \lesssim 2$ it covers one or two periods over the wavenumbers that the CMB measures and resembles a running of the spectral index \cite{Kobayashi:2010pz,Meerburg:2014bpa}.

The same resonance produces a bispectrum with an amplitude \cite{Flauger:2010ja}
\begin{equation}
    f_\mathrm{NL}^\mathrm{res} = \frac{3\sqrt{2\pi}\, b}{8 (f\phi_*)^{3/2}} = \frac{\delta n_\mathrm{s}\, \omega^2}{8},
    \label{eq:fnl}
\end{equation}
so that the two signals are related. At our best fit $f_\mathrm{NL}^\mathrm{res} \approx 8$. The Planck 2018 bispectrum shows no significant resonant signal for $0 < \omega < 1000$ \cite{Planck:2019kim}. However, its sensitivity is far too low to test such an amplitude: already at $\omega \approx 33$ the error of $f_\mathrm{NL}^\mathrm{res}$, which is set by cosmic variance, is about 80 \cite{Planck:2015sxf}, and the errors grow with frequency \cite{Planck:2019kim}. In a two-field extension of the model the modulation also excites heavy fields, such as the moduli of the compactification, when its frequency exceeds their masses, and these fields then leave signals of their own in the bispectrum \cite{Pajer:2024ckd}. At much higher frequencies than those considered here, with 350 to 1000 oscillations per decade in $k$, i.e.\ $\omega \approx 950$--2700, the signal of a small modulation lies mainly in correlation functions of higher order than the power spectrum, of which those with up to about nine points can be observable \cite{Creminelli:2025tae}. In this work we use only the power spectrum.

\subsection{The logarithmic oscillation template}
\label{sec:theory:template}

We parametrise the primordial spectrum as
\begin{equation}
    P(k) = A_\mathrm{s} \left(\frac{k}{\kstar}\right)^{n_\mathrm{s}-1+\frac{1}{2}\nrun\ln(k/\kstar)} \left[ 1 + \Aosc \sin\left( \omega \ln\frac{k}{\kstar} + \phi \right) \right],
    \qquad \kstar = 0.05\,\mathrm{Mpc}^{-1},
    \label{eq:template}
\end{equation}
which is eq.~\eqref{eq:monodromy_spectrum} with $\Aosc = \delta n_\mathrm{s}$ and a running $\nrun = \dd n_\mathrm{s}/\dd\ln k$ of the smooth spectrum. The sign of the frequency, the cosine and the phase $\Delta\varphi$ are absorbed into the phase $\phi$. Without the running this is the template of the Planck analyses \cite{Planck:2015sxf,Planck:2018jri}, which write it with a cosine, i.e.\ with $\phi$ shifted by $\pi/2$.

The running is included because of the low frequencies. Expanding the oscillation around the pivot scale, the logarithm of the spectrum changes by $\Aosc\sin\phi + \Aosc\,\omega\cos\phi\, x - \frac{1}{2}\Aosc\,\omega^2 \sin\phi\, x^2$ with $x = \ln(k/\kstar)$. At $\omega \lesssim 2$, where the CMB sees only one or two periods, the oscillation changes the amplitude, the tilt and the running of the spectrum, by $\Aosc\,\omega\cos\phi$ in $n_\mathrm{s}$ and by $-\Aosc\,\omega^2\sin\phi$ in $\nrun$. In a model without running a slow oscillation is the only source of curvature of the spectrum, and a preference for one could be mistaken for a preference for the oscillation. With $\nrun$ free and a prior wide enough for the running to cancel the oscillation's own, the oscillation must instead improve the fit beyond what a smooth running does.

We sample $\omega$ uniformly in $[1, 200]$. The Planck analyses sample $\log_{10}\omega_\mathrm{log} \in [0, 2.1]$, i.e.\ $\omega \approx 1$--126 \cite{Planck:2018jri}, as do most recent analyses of CMB data \cite{Zeng:2018ufm,Antony:2024vrx,Raffaelli:2026bhk}, and some stop at $\omega = 100$ \cite{Planck:2013jfk,Peng:2025vda,Nerval:2026iev}. Higher frequencies have been searched in the Planck 2013 temperature data, up to $\omega \approx 250$ \cite{Meerburg:2013dla,Easther:2013kla} and, with a drifting frequency, up to $\omega \approx 1000$ \cite{Flauger:2014ana}. In the clustering of BOSS and eBOSS galaxies the search extends to $\omega = 360$ \cite{Mergulhao:2023ukp}. These searches found no significant feature. A prior uniform in $\omega$ weights all frequencies equally, whereas one uniform in $\log\omega$ would favour the low frequencies \cite{Hamann:2021eyw}, and since the marginalised limit on $\Aosc$ depends on the frequency range \cite{Calderon:2025xod}, it should be compared with limits over the same range. Our upper bound lies at the frequency above which the effective field theory of the oscillating background becomes strongly coupled (Sec.~\ref{sec:theory:monodromy}). A scan of the likelihood, in which the improvement in $\chi^2$ is maximised over $\Aosc$ and $\phi$ on a grid in $\omega$ with the other parameters fixed at a reference point, shows that the data has maxima at high frequency as well. For \PDP{} the highest maximum lies at $\omega \simeq 51.5$, where the oscillation lowers $\chi^2$ by 10.2, and above $\omega = 60$ the largest improvements are 8.6 at $\omega \simeq 196$, 8.0 at 179 and 6.5 at 100.75. For \HWC{} the largest improvement above $\omega = 60$ is 6.2 at $\omega \simeq 157.75$. These maxima at high frequency require amplitudes up to $\Aosc \approx 0.05$. With the Planck PR3 likelihoods ref.~\cite{Raffaelli:2026bhk} finds a further maximum at $\omega \approx 65$, which lowers $\chi^2$ by 7 to 9, but not with the CamSpec PR4 likelihood used here.

The phase is periodic, and the likelihood depends on $\phi$ only modulo $2\pi$, so that any window of width $2\pi$ describes the same model. However, a sampler, or an emulator trained in a box, sees the boundaries of the window, and a likelihood peak cut by a boundary is harder both to sample and to learn. We therefore place the window $[\phi_0 - \pi, \phi_0 + \pi]$ so that it cuts the peaks of the likelihood as little as possible. To find the peaks we fit the oscillation on a grid in $\omega$ at a reference point of the other parameters. Since the CMB spectra depend almost linearly on $\Aosc\cos\phi$ and $\Aosc\sin\phi$ when the other parameters are fixed, a few amplitudes and phases per frequency suffice to locate the peaks. This gives $\phi_0 = 0.42$ for \PDP{} and $\phi_0 = 0.48$ for \HWC{}.

\section{Numerical setup}
\label{sec:numerics}

\subsection{\CLASSpp{} and the primordial spectrum}
\label{sec:numerics:classpp}

The CMB spectra are computed with \CLASSpp{}, a \CC{} version of \class{} \cite{Lesgourgues:2011re,Blas:2011rf}. The primordial spectrum of eq.~\eqref{eq:template} is passed to it as an external table with 1000 points per decade in $k$, so that even at $\omega = 200$ the phase of the oscillation advances by less than 0.5 between neighbouring points. The non-linear correction to the lensing potential is computed with halofit \cite{Takahashi:2012em}. The neutrinos are massless, with $N_\mathrm{eff} = 3.044$. The neutrino mass is not degenerate with the oscillation, and since we are interested in the improvement of the fit due to the oscillation, the small change of the featureless model which this implies is not important. We use the public version of \CLASSpp{} (\url{https://github.com/AarhusCosmology/CLASSpp_public}) at the commit 0a57bbc6 of its master branch, of 27 October 2026, without modifications.

The lensed CMB spectra are computed at a subset of multipoles and interpolated in $\ell$, and at high frequency this interpolation smooths the oscillation in $\ell$, whose period is $\Delta\ell \approx 2\pi\ell/\omega$. With the default sampling of \CLASSpp{} the improvement in $\chi^2$ due to the oscillation is very poorly reproduced at high frequency, and at $\omega \simeq 50$ it was underestimated by more than an order of magnitude in a preliminary scan with the Planck data. We therefore use a much denser sampling in $\ell$, with a logarithmic step of 1.013 instead of 1.12 and a linear step of 8 instead of 50. At high frequency the transfer functions must also be sampled more densely in $k$, and we use three and four times smaller steps than the defaults (\texttt{q\_linstep\_trans}~$= 0.15$ and \texttt{q\_logstep\_trans}~$= 90$), together with the accurate computation of the lensed spectra. Compared with a converged sampling, this changes the improvement in $\chi^2$ at the maxima of the likelihood by at most 0.09 for $\omega < 200$ with \PDP{}. A sampling that is denser only in $\ell$ is sufficient up to $\omega \approx 90$, but above this frequency the improvements due to the oscillation can then be wrong by up to 10 in $\chi^2$ (Appendix~\ref{app:classpp}). One evaluation of the likelihood then takes about 24 core seconds.

\subsection{Cosmological model: parameters and priors}
\label{sec:numerics:params}

The sampled cosmological parameters are
\begin{equation}
    \theta = \{\, \omega_\mathrm{b},\, \omega_\mathrm{cdm},\, 100\,\theta_\mathrm{s},\, \ln(10^{10}A_\mathrm{s}),\, n_\mathrm{s},\, \nrun,\, \tau_\mathrm{reio},\, \Aosc,\, \omega,\, \phi \,\},
\end{equation}
i.e.\ the six parameters of $\Lambda$CDM, the running of the spectral index and the three parameters of the oscillation. We sample the angular scale of the sound horizon at recombination, $\theta_\mathrm{s}$, rather than $H_0$, which is a derived parameter. The priors are uniform, with the ranges listed in Table~\ref{tab:priors}. For the $\Lambda$CDM parameters they extend to about ten standard deviations of the Planck constraints on either side of the mean, and the range of $n_\mathrm{s}$ is wider still because at low frequency the oscillation shifts the tilt. The range of the running, $|\nrun| \leq 0.1$, is wide enough for the running to cancel the curvature of a slow oscillation, $\Aosc\,\omega^2 \approx 0.1$ at $\omega \approx 2$ for amplitudes that the data allows there. The amplitude has a uniform prior $0 \leq \Aosc \leq 0.1$. Larger amplitudes are disfavoured by the data: at $\Aosc = 0.1$ the profile likelihood of \PDP{} is worse than the best fit by $\Delta\chi^2 = 10.6$ (Appendix~\ref{app:profiles}), and its maximum there lies at high frequency, where the imprint of the oscillation on the spectra is smallest (Sec.~\ref{sec:results:profiles}).

In addition to the cosmological parameters we sample the nine nuisance parameters of the CamSpec likelihood, i.e.\ the overall calibration $A_\mathrm{Planck}$ (shared with the lensing likelihood) and the polarisation calibrations $c_\mathrm{TE}$ and $c_\mathrm{EE}$ with Gaussian priors, as well as the amplitudes and slopes of the power-law foreground model at 143, 217 and $143\times217$\,GHz with uniform priors. In total we sample 19 parameters for \PDP{} and 22 for \HWC{}.

\begin{table}[tb]
\centering
\renewcommand{\arraystretch}{1.1}
\begin{tabular}{rlccc}
\cmidrule[\heavyrulewidth](lr){2-5}
 & \textbf{Parameter} & \textbf{Prior} & \multicolumn{2}{c}{\textbf{Training boxes}} \\
 & & & \PDP{} & \HWC{} \\
\cmidrule(lr){2-5}
$\Lambda$CDM & $\omega_\mathrm{b}$ & $[0.021, 0.0238]$ & prior & prior \\
 & $\omega_\mathrm{cdm}$ & $[0.11, 0.129]$ & prior & prior \\
 & $100\,\theta_\mathrm{s}$ & $[1.039, 1.0448]$ & prior & prior \\
 & $\ln(10^{10}A_\mathrm{s})$ & $[2.9, 3.2]$ & prior & prior \\
 & $n_\mathrm{s}$ & $[0.92, 1.02]$ & prior & prior \\
 & $\mathrm{d}n_\mathrm{s}/\mathrm{d}\ln k$ & $[-0.1, 0.1]$ & prior & prior \\
 & $\tau_\mathrm{reio}$ & $[0.01, 0.12]$ & prior & prior \\
\cmidrule(lr){2-5}
Oscillation & $A_\mathrm{osc}$ & $[0, 0.1]$ & prior & prior \\
 & $\omega$ & $[1, 200]$ & Table~\ref{tab:zones} & Table~\ref{tab:zones} \\
 & $\phi$ & $[\phi_0 - \pi, \phi_0 + \pi]$ & $[-2.719, 3.564]$ & $[-2.663, 3.620]$ \\
\cmidrule[\heavyrulewidth](lr){2-5}
\end{tabular}
\caption{Uniform priors on the cosmological parameters, and the training boxes of the emulators where they differ from the prior. The prior volume is split in $\omega$ between several emulators for each dataset (Sec.~\ref{sec:numerics:client}), whose boxes in $\omega$ are given in Table~\ref{tab:zones} in Appendix~\ref{app:split}. The phase window has width $2\pi$ and is placed so that it cuts the peaks of the likelihood as little as possible (Sec.~\ref{sec:theory:template}). For $\Aosc$ the runs used the prior $0 \leq \Aosc \leq 0.5$ in \cobaya{}, the emulators were trained on $0 \leq \Aosc \leq 0.1$, and all results are given for the prior $0 \leq \Aosc \leq 0.1$.}
\label{tab:priors}
\renewcommand{\arraystretch}{1.0}
\end{table}

\subsection{Datasets}
\label{sec:numerics:data}

We use the following likelihoods, all as implemented in \cobaya{} \cite{Torrado:2020dgo}:
\begin{itemize}
    \item \textbf{Planck}: the Planck 2018 low-$\ell$ temperature (Commander) and $E$-mode polarisation (SimAll) likelihoods \cite{Planck:2019nip}, the high-$\ell$ TT, TE and EE CamSpec likelihood built on the Planck PR4 (NPIPE) maps \cite{Rosenberg:2022sdy}, and the Planck PR4 CMB lensing reconstruction \cite{Carron:2022eyg}.
    \item \textbf{DESI BAO}: the baryon acoustic oscillation measurements of the second data release (DR2) of the Dark Energy Spectroscopic Instrument, from all tracers \cite{DESI:2025zgx}.
    \item \textbf{Pantheon+}: the Pantheon+ compilation of type Ia supernovae \cite{Brout:2022vxf}, uncalibrated, with the absolute magnitude marginalised analytically.
\end{itemize}
We refer to this dataset as Planck+DESI+Pantheon+ (\PDP{}). The same dataset and the more aggressive combination \HWC{} (Sec.~\ref{sec:actspt}) are also used in ref.~\cite{Uggerhoj:2026jet}, in an analysis of self-interacting neutrinos with \client{}.

At $\omega = 200$ the period of the oscillation in $\ell$ is about 30 multipoles at $\ell \approx 1000$, comparable to the width of the bins of binned likelihoods, which therefore smooth out such oscillations \cite{Raffaelli:2026bhk}. However, the CamSpec likelihood compares the spectra multipole by multipole, with a covariance that includes the correlations between neighbouring multipoles, so that its sensitivity is not reduced by binning.

\subsection{Likelihood emulation with \client{}}
\label{sec:numerics:client}

A Monte Carlo analysis of this model calling \CLASSpp{} directly is not feasible (Appendix~\ref{app:cost}). We therefore use \client{} \cite{Janken:2025wlq} (\clienturl{}), which emulates the likelihood itself. A neural network learns the log-posterior of the data (the sum of the log-likelihoods and the log-prior) as a function of all sampled parameters, cosmological and nuisance, inside a training box. We write $\mathcal{P}$ for the product of the likelihoods and the Gaussian priors on the calibration parameters of the CMB likelihoods (Secs.~\ref{sec:numerics:params} and \ref{sec:actspt}), which we regard as part of the likelihood, and all other priors are uniform. We give the errors of the emulator and the thresholds of its training as differences in $\ln\mathcal{P}$, and all comparisons of fits in $\Delta\chi^2 = -2\Delta\ln\mathcal{P}$. A difference of 1 in $\ln\mathcal{P}$ is therefore a difference of 2 in $\chi^2$. The training points are selected by the code in two stages, and the likelihood at them is computed with \CLASSpp{} and \cobaya{}. The settings are listed in Tables~\ref{tab:client} and \ref{tab:client_b}, and Appendix~\ref{app:client} gives the details.

In the first stage the emulator is trained iteratively, starting from 3000 points of a Latin hypercube. In each iteration a network is trained on all points computed so far and sampled at a temperature $T = 7$, with its posterior raised to the power $1/T$, and the true likelihood is computed at 3000 points of this tempered chain. Before the chain is run, an audit computes the true likelihood at the maxima of the network which lie well above any point computed so far. A maximum which is not present in the true likelihood is then removed by the next training. The first stage ends when the network predicts the true likelihood at its new tempered points with a scatter below $1.2\,T$ in $\ln\mathcal{P}$ for two consecutive networks.

In the second stage the emulator is refined in ten boosts. Before a boost \client{} identifies the modes of the posterior of the current network, i.e.\ the regions of high probability between which a Markov chain at $T = 1$ does not move freely (Appendix~\ref{app:boosts}). Peaks in $\omega$ which the chain moves between belong to the same mode, and all but one of our emulators have only one mode. The boost then computes the true likelihood at 10\,000 points drawn from this posterior at $T = 1$, with the same weight for all modes, so that a mode which holds only a small part of the posterior probability is learned as well as the largest one. In the boosts each network is trained 12 times, and the true likelihood at 100 samples of the posterior of each of them decides which one is kept. The emulator is used only inside a region made of ellipsoids around the boost points, and outside this region its value is replaced by a large penalty. Samplers and optimisers therefore do not find maxima of the network far from the training points. For \PDP{} the posterior after four boosts is already indistinguishable from the final one (Appendix~\ref{app:boosts}). We nevertheless continued to ten boosts, in order to demonstrate that the pointwise error of the emulators can be driven very low.

We split the prior volume in $\omega$ and train separate emulators for the parts, with overlapping training boxes (Table~\ref{tab:zones} in Appendix~\ref{app:split}). The lowest part covers the low frequencies, at which the oscillation spans only a few periods over the scales that the CMB measures and is degenerate with $n_\mathrm{s}$ and $\nrun$ along a broad ridge. Above it, we performed the domain splitting such that each emulator trains on a roughly comparable number of local maxima of the likelihood. Choosing the parts, and the bands in which neighbouring parts are joined, is the only step that is done by hand. Each part is an ordinary run of \client{} with the same likelihood and the same prior, and after the training one command of \client{} checks that the emulators of a dataset are compatible and combines them into a single network. In this network the emulators are used only inside their own parts, and neighbouring emulators are blended smoothly in a narrow band inside the overlap of their training boxes. The combined emulator is sampled, integrated and maximised with the same tools as an emulator trained on the whole prior volume, and all results below are obtained from it. Each emulator was trained on 96 cores, and together the emulators of \PDP{} were trained on \ensuremath{9.2\times10^{5}} likelihood evaluations.

\subsection{Parameter inference and model comparison}
\label{sec:numerics:inference}

The posterior distributions are sampled from the combined emulator at $T = 1$ with an affine invariant ensemble sampler \cite{Goodman:2010dyf,Foreman-Mackey:2012any}. The region of the combined emulator consists of the ellipsoids of its parts, and a single ensemble sampler does not move reliably between separated parts of a region. \client{} therefore samples the parts separately and weights them by their evidences (85\,504 samples in total). The evidence of a part is estimated from its own chain. A mixture of Gaussians is fitted to half of the samples, and the evidence follows from the ratio of the emulated posterior to this normalised density, evaluated at the other samples and at draws from the mixture, a method known as bridge sampling \cite{Meng:1996}. The derived parameters, such as $H_0$ and $\sigma_8$, are computed at the samples with a second, smaller network, which \client{} trains alongside the emulator on the derived parameters returned by \cobaya{} at the training points. The chains are analysed using GetDist \cite{Lewis:2019xzd}. To test the posterior we compute the true likelihood at 500 of its samples and weight each by $\e^{\ln\mathcal{P}_\mathrm{true} - \ln\mathcal{P}_\mathrm{emu}}$ (Appendix~\ref{app:accuracy}).

The Bayesian evidence is computed by nested sampling of the combined emulator with BEST \cite{Nygaard:2026fgl,Nygaard:2026hbr}, with 2048 live points of which 128 are replaced per iteration. Since the network has learned the log-posterior including the normalised prior densities, its integral over the training box is the evidence, provided that the box contains the posterior. We repeat the nested sampling with four different seeds and quote the mean and its standard error, since the scatter between the runs, 0.21, is larger than the error of 0.14 which a single run quotes. This error does not include the error of the emulator, which we estimate from the comparison with the true likelihood (Appendix~\ref{app:accuracy}). The evidence of the featureless model, $\Lambda$CDM with running, is computed from the same emulator with the prior on $\Aosc$ narrowed to $[0, 10^{-4}]$. The likelihood then no longer depends on $\omega$ and $\phi$, and the result is the evidence of the model without the oscillation, with the same priors on all other parameters. As a check of the Bayes factor we also compute the Savage--Dickey density ratio \cite{Trotta:2005ar}, i.e.\ the marginalised posterior density at $\Aosc = 0$ divided by the prior density, from the posterior samples of the emulator.

The profile likelihoods are calculated with the profile tool of \client{}, and their values are those of the true likelihood. For each pair of oscillation parameters, $(\Aosc, \omega)$, $(\Aosc, \phi)$ and $(\omega, \phi)$, we fix the pair at the nodes of a $50\times50$ grid and maximise over all other parameters, i.e.\ over 17 parameters for \PDP{} and 20 for \HWC{}. This is done in two steps. In the first step the emulator is maximised, at all nodes together. Local maximisations from many starting points are combined with random steps and with starts at the optima of the neighbouring nodes, until no node improves any further. The second step begins only if this scan has converged and no optimum lies at the boundary of the region in which the emulator is used. In it each node is refined with the true likelihood, starting at the optimum of the emulator at that node. Let $\psi$ denote the parameters that are maximised at a node and $\psi_0$ the best point found so far. Around $\psi_0$ the code models the true log-posterior as
\begin{equation}
    \ln\mathcal{P}_\mathrm{true}(\psi) \simeq \ln\mathcal{P}_\mathrm{emu}(\psi) + \epsilon(\psi_0) + g\cdot(\psi - \psi_0) + \frac{1}{2}\,(\psi - \psi_0)^\mathrm{T}\, Q\, (\psi - \psi_0),
    \label{eq:profile_model}
\end{equation}
where $\epsilon = \ln\mathcal{P}_\mathrm{true} - \ln\mathcal{P}_\mathrm{emu}$ is the error of the emulator. Near the optimum $\epsilon$ varies slowly compared with the emulator itself, so that a correction of low order is sufficient. Its gradient $g$ is measured by finite differences of the true likelihood, and the matrix $Q$, which corrects the curvature of the emulator, is updated from the steps taken. In each step the right-hand side of eq.~\eqref{eq:profile_model} is maximised within a trust region around $\psi_0$, and the true likelihood is computed at the new point, which replaces $\psi_0$ if it is better. The refinement of a node ends when the model predicts no further gain, or after 150 evaluations of the true likelihood. Along the rows of the grid $g$ and $Q$ are carried over from one node to the next. The value of a node is therefore always a value of the true likelihood, and the best fit is finally refined with all parameters free (Appendix~\ref{app:profiles}). The one-dimensional profile of a parameter is the best value over the other parameter of the surfaces that contain it. Since all priors except the Gaussian priors on the calibration parameters are uniform (Sec.~\ref{sec:numerics:client}), the best fits and the profile likelihoods maximise the posterior.

To compare the oscillation model with a featureless spectrum we use the profile at $\Aosc = 0$, where $\omega$ and $\phi$ have no effect, i.e.\ $\Lambda$CDM with a running of the spectral index. We quote $\Delta\chi^2$ between the best fits and the difference in the Akaike information criterion \cite{Akaike:1974vps}, $\Delta\mathrm{AIC} = \Delta\chi^2 + 2\Delta N$, with $\Delta N = 3$. Note that $\omega$ and $\phi$ are undefined at $\Aosc = 0$, so that neither $\Delta\chi^2$ nor $\Delta\mathrm{AIC}$ follows its usual distribution, and a significance requires a look-elsewhere correction \cite{Gross:2010qma,Bayer:2020pva}.

\section{Results}
\label{sec:results}

We first present the results for \PDP{} (Sec.~\ref{sec:results:pp}), and in Sec.~\ref{sec:actspt} we add the small-scale CMB data from ACT and SPT.

\subsection{Planck, DESI BAO and Pantheon+ (\PDP{})}
\label{sec:results:pp}

\subsubsection{Parameter inference}
\label{sec:results:params}

\begin{figure}[tb]
\centering
\includegraphics[width=\textwidth]{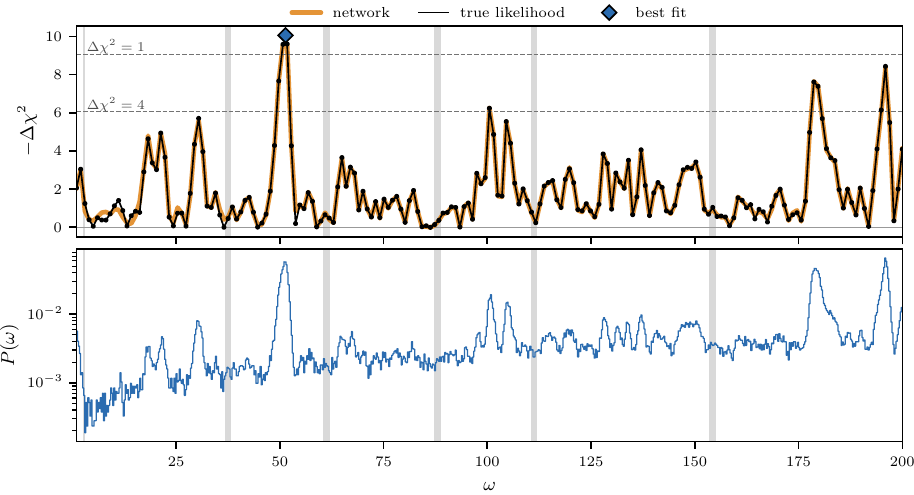}
\caption{The frequency of the oscillation for \PDP{}. Top: the improvement in $\chi^2$ relative to the best fit without oscillation, from the true profile likelihood in $(\omega, \phi)$, i.e.\ with $\Aosc$ and all other parameters maximised with the true likelihood (black), and the same profile calculated with the emulator (orange), relative to the same best fit without oscillation. The diamond is the best fit. It is found in the plane $(\Aosc, \phi)$, in which $\omega$ is maximised, and lies between two nodes of the grid in $\omega$. The spacing of the grid, about 1 in $\omega$, is comparable with the width of the maxima in $\omega$, so that the profile on the grid lies slightly below the best fit. The dashed lines show $\Delta\chi^2 = 1$ and 4 relative to the best fit. Bottom: the marginalised posterior density of $\omega$ on a logarithmic scale. Neighbouring emulators are blended in the grey bands.}
\label{fig:omega}
\end{figure}

With frequencies up to $\omega = 200$ the posterior of \PDP{} is spread over many narrow peaks in $\omega$ (Fig.~\ref{fig:omega}). The peak of the best fit, at $\omega \approx 51$, contains only 12\% of the posterior, and 78\% lies above $\omega = 60$. The reasons are that the likelihood has maxima of almost the same height at high frequency, that the data allows a wider range of amplitudes there (Sec.~\ref{sec:results:profiles}) and that frequencies above $\omega = 60$ make up 70\% of the prior. The posterior therefore does not single out any frequency.

\begin{table}[tb]
\centering
\renewcommand{\arraystretch}{1.1}
\begin{tabular}{rlcc}
\cmidrule[\heavyrulewidth](lr){2-4}
& \textbf{Parameter} & \textbf{\PDP{}} & \textbf{\HWC{}} \\
\cmidrule(lr){2-4}
$\Lambda$CDM & $\omega_\mathrm{b}$ & $0.02230\pm 0.00012$ & $0.02239\pm 0.00010$ \\
 & $\omega_\mathrm{cdm}$ & $0.11833\pm 0.00061$ & $0.11822\pm 0.00058$ \\
 & $100\,\theta_\mathrm{s}$ & $1.04184\pm 0.00023$ & $1.04170\pm 0.00022$ \\
 & $\ln(10^{10}A_\mathrm{s})$ & $3.040\pm 0.014$ & $3.055\pm 0.011$ \\
 & $n_\mathrm{s}$ & $0.9658\pm 0.0039$ & $0.9725\pm 0.0030$ \\
 & $\mathrm{d}n_\mathrm{s}/\mathrm{d}\ln k$ & $-0.0034\pm 0.0066$ & $0.0055\pm 0.0051$ \\
 & $\tau_\mathrm{reio}$ & $0.0551\pm 0.0069$ & $0.0620\pm 0.0057$ \\
\cmidrule(lr){2-4}
Oscillation & $A_\mathrm{osc}$ & $< 0.056$ & $< 0.062$ \\
 & $\omega$ & multimodal & multimodal \\
 & $\phi$ & multimodal & multimodal \\
\cmidrule(lr){2-4}
Derived & $H_0\;[\mathrm{km\,s^{-1}\,Mpc^{-1}}]$ & $68.38\pm 0.28$ & $68.46\pm 0.25$ \\
 & $\sigma_8$ & $0.8166\pm 0.0058$ & $0.8261\pm 0.0042$ \\
 & $S_8$ & $0.8176\pm 0.0078$ & $0.8261\pm 0.0063$ \\
 & $\Omega_\mathrm{m}$ & $0.3007\pm 0.0035$ & $0.3000\pm 0.0032$ \\
 & $r_\mathrm{drag}\;[\mathrm{Mpc}]$ & $147.64\pm 0.19$ & $147.56\pm 0.18$ \\
\cmidrule[\heavyrulewidth](lr){2-4}
\end{tabular}
\caption{Marginalised constraints: mean and 68\% interval, or the 95\% upper limit. The posterior is multimodal in $\omega$ and $\phi$, and its peaks are discussed in the text (Figs.~\ref{fig:omega} and \ref{fig:omega_wc}).}
\label{tab:params}
\renewcommand{\arraystretch}{1.0}
\end{table}

\begin{figure}[tb]
\centering
\includegraphics[width=\textwidth]{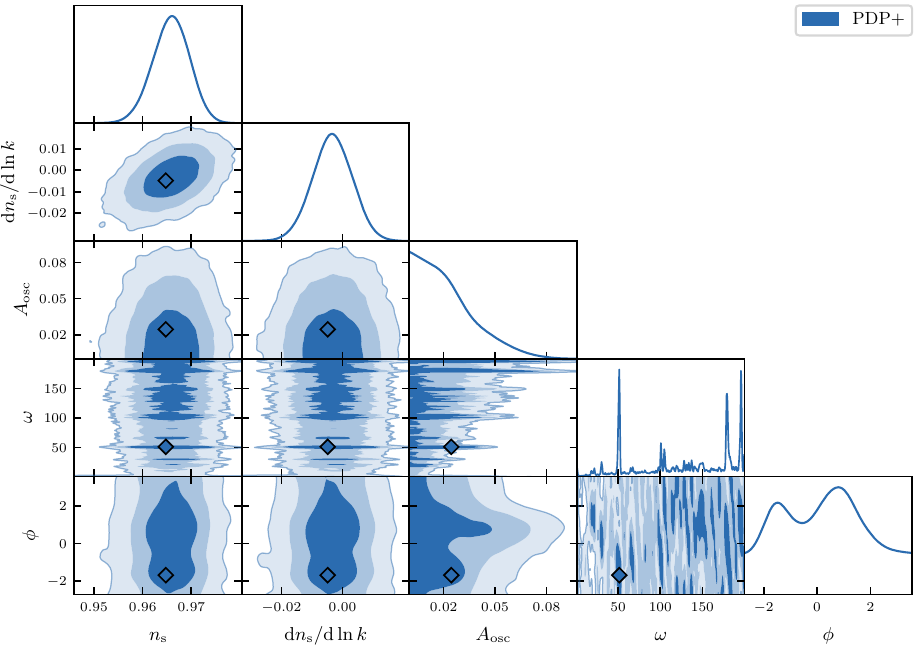}
\caption{The posterior of the spectral index, the running and the oscillation parameters for \PDP{} (68\%, 95\% and 99.7\% regions). The diamond is the best fit of the true profile likelihood. The panels with $\omega$ are drawn from histograms in bins of $\Delta\omega = 0.5$, which resolve the many narrow peaks of its posterior.}
\label{fig:triangle}
\end{figure}

The marginalised constraints are listed in Table~\ref{tab:params}, and Fig.~\ref{fig:triangle} shows the posterior of the spectral index, the running and the oscillation parameters. The posteriors of the $\Lambda$CDM parameters are very close to Gaussian, and the running is consistent with zero. The amplitude of the oscillation is consistent with zero as well, and we find $\Aosc < 0.056$ at 95\%~CL. This limit is set by the high frequencies, where the data allows larger amplitudes (Sec.~\ref{sec:results:profiles}), and for $\omega \leq 60$ alone it is $\Aosc < 0.031$. This is comparable with the limits of 2--3\% in the recent literature, although these are obtained for other ranges of the frequency, and partly with a prior uniform in $\log\omega$ \cite{Planck:2018jri,Raffaelli:2026bhk,Ballardini:2022wzu}. At low frequency, where 0.5\% of the posterior lies, the oscillation spans one or two periods over the measured wavenumbers and can be compensated by a change of $n_\mathrm{s}$ and the running (Sec.~\ref{sec:theory:template}).

\subsubsection{Profile likelihoods}
\label{sec:results:profiles}

The profile likelihood of \PDP{} has a fairly well-defined best fit at $\omega = 51.4$, $\Aosc = 0.024$ and $\phi = -1.69$, which lowers $\chi^2$ by 10.1 relative to the best fit without oscillation. The peak is narrow (Fig.~\ref{fig:omega}). However, the maxima at $\omega \approx 196$ and 179 are worse than the best fit by only 1.6 and 2.3 in $\chi^2$. The data therefore does not single out the frequency of the best fit (Sec.~\ref{sec:results:comparison}). The other maxima and the best fit without oscillation are discussed in Appendix~\ref{app:profiles}.

The frequency of the best fit also differs from those of the Planck 2018 analyses, which found $\omega \approx 30$ with temperature data and $\omega \approx 18$ with polarisation \cite{Planck:2018jri}. \PDP{} has maxima at these frequencies too, but lower ones (Fig.~\ref{fig:omega}), and which frequency ranks highest is known to depend on the likelihood used \cite{Raffaelli:2026bhk}. With the CamSpec PR4 likelihood, ref.~\cite{Raffaelli:2026bhk} finds a best fit at the same frequency, with the same amplitude and the same improvement in $\chi^2$ as ours, and a global significance of $1.3\sigma$ after the look-elsewhere correction.

\begin{figure}[tb]
\centering
\includegraphics[width=\textwidth]{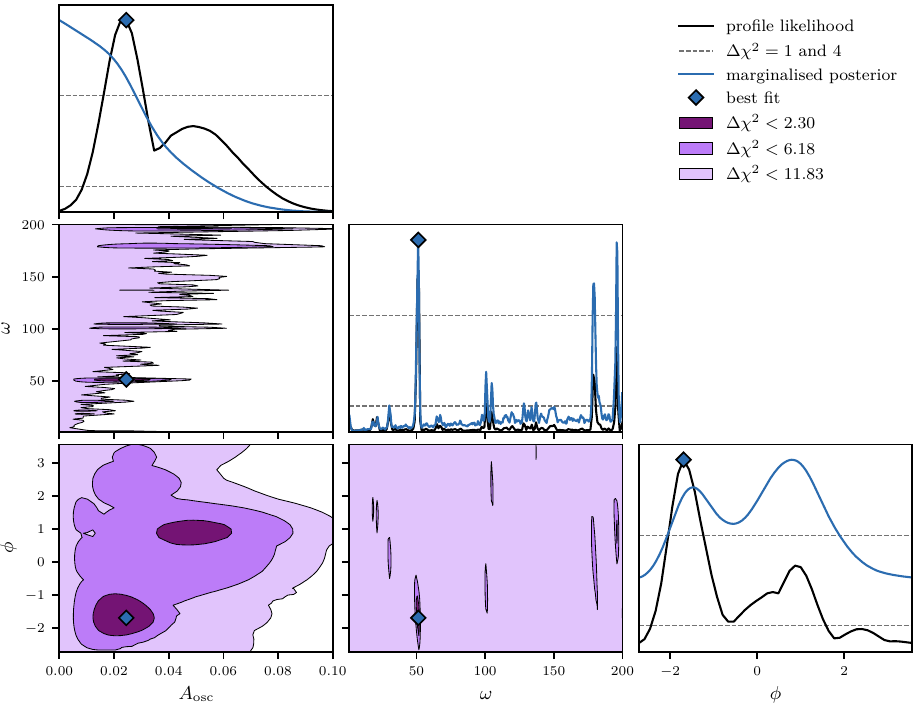}
\caption{The profile likelihood of \PDP{} in the oscillation parameters. Below the diagonal: the two-dimensional profiles. The planes $(\Aosc, \omega)$ and $(\omega, \phi)$ are each calculated in four windows in $\omega$ of $50\times50$ nodes, joined into one surface, and the plane $(\Aosc, \phi)$ on $50\times50$ nodes. At each node the third oscillation parameter and all other parameters are maximised with the true likelihood. In the two lowest windows of $(\Aosc, \omega)$ the grid ends at $\Aosc = 0.06$ and 0.09, respectively (white above). The regions with $\Delta\chi^2 < 2.30$, 6.18 and 11.83 relative to the best fit (diamond), i.e.\ the Gaussian 68\%, 95\% and 99.7\% levels for two parameters, are shaded. In the $(\omega, \phi)$ plane no node is worse than the best fit without oscillation, $\Delta\chi^2 = 10.1$, since $\Aosc = 0$ is allowed at all nodes, and the whole plane therefore lies inside the 99.7\% level. Diagonal: the one-dimensional profile likelihood ratio $\e^{-\Delta\chi^2/2}$ relative to the best fit (black), from the better of the two surfaces that contain the parameter, the best fit (diamond) and the marginalised posterior with a maximum of 1 (blue). In $\omega$ the best fit lies between two nodes of the grid, and the profile is therefore below 1 (Fig.~\ref{fig:omega}). For $\omega$ the posterior is a histogram in bins of width 0.5. The dashed lines show $\Delta\chi^2 = 1$ and 4 for the profile.}
\label{fig:profile2d}
\end{figure}

Figure~\ref{fig:profile2d} shows the profile likelihood in the three planes of the oscillation parameters. In the plane of the frequency and the phase the peaks are tilted, and at the main peak the phase of the best fit changes with the frequency as $\dd\phi/\dd\omega \approx -0.27$. This means that a change of frequency is compensated by a change of phase which keeps the oscillation in place at the wavenumber $k = \kstar\,\e^{-\dd\phi/\dd\omega} \approx 0.065\,\mathrm{Mpc}^{-1}$, which projects onto multipoles $\ell \approx 900$.

Since \client{} gives both the profile likelihood and the posterior, the two can be compared directly, and the diagonal of Fig.~\ref{fig:profile2d} shows them together. For the frequency and the phase their peaks lie at the same values. However, for the amplitude they differ markedly. The profile peaks at $\Aosc \approx 0.024$ and disfavours $\Aosc = 0$ by $\Delta\chi^2 = 10.1$, but the posterior is largest at $\Aosc = 0$. The reason is that at most frequencies and phases the data allows only a small amplitude, and after marginalisation this large volume at small amplitude dominates the posterior of $\Aosc$.

At large amplitude the two differ in the opposite way. There the best fit moves to the maximum at $\omega \approx 196$, and the profile stays within $\Delta\chi^2 = 4$ of the best fit up to $\Aosc = 0.075$. At high frequency the imprint of a given amplitude on the spectra is smaller, since the projection onto multipoles averages over a fast oscillation in $\ln k$, and from $\omega = 60$ to 200 it decreases by more than a factor of three (Table~\ref{tab:classpp_cl}). However, the data allows a large amplitude only near the narrow maxima at high frequency, and the posterior therefore gives the stronger upper limit, $\Aosc < 0.056$. Both the profile and the posterior are needed to describe the constraints on the amplitude.

The weaker imprint at high frequency also affects the posterior of $\omega$. Compared with the profile, the posterior rises towards high frequencies, and also towards $\omega \approx 1$ (Figs.~\ref{fig:omega} and \ref{fig:profile2d}). The reason is again the volume after marginalisation: at each frequency the posterior integrates over the amplitudes that the data allows, and their range depends on the frequency. At fixed $\omega$ the 95\% upper limit on $\Aosc$ is about 0.006 for $3 \leq \omega < 10$, but about 0.07 for $\omega \geq 180$ and 0.04 for $\omega < 2$, where the oscillation is degenerate with $n_\mathrm{s}$ and the running.

\subsubsection{Model comparison}
\label{sec:results:comparison}

Table~\ref{tab:model_comparison} summarises the comparison with a featureless spectrum. With \PDP{} the oscillation lowers $\chi^2$ by 10.1 at the cost of three additional parameters, i.e.\ $\Delta\mathrm{AIC} = -4.1$. This is a mild preference for the oscillation in terms of the AIC. However, fits of the template to featureless simulated skies typically lower $\chi^2$ by roughly 10 over a wide range of frequencies \cite{Easther:2013kla,Meerburg:2013cla}, and the profile of \PDP{} itself has maxima of almost the same height at high frequency (Sec.~\ref{sec:results:profiles}).

We have therefore estimated the global significance from the profile likelihood in $\omega$, following ref.~\cite{Gross:2010qma}. Without an oscillation the improvement in $\chi^2$ at fixed $\omega$ approximately follows a $\chi^2$ distribution with two degrees of freedom, the amplitude and the phase, so that the local $p$-value of the best fit is 0.007. However, the improvement exceeds 1 in 30 separate ranges of $\omega$ between $\omega = 1$ and 200. From the number of such ranges, the expected number of maxima as high as the best fit without any oscillation is 1.0 (Appendix~\ref{app:profiles}). The global significance of the best fit is therefore negligible.

\begin{table}[tb]
\centering
\renewcommand{\arraystretch}{1.1}
\begin{tabular}{rlcccc}
\cmidrule[\heavyrulewidth](lr){2-6}
 & \textbf{Model} & $N$ & $\Delta\chi^2$ & $\Delta$AIC & $\ln B$ \\
\cmidrule(lr){2-6}
\PDP{} & $\Lambda$CDM + running & 7 & 0 & 0 & 0 \\
 & + oscillation & 10 & $-10.1$ & $-4.1$ & $-1.0 \pm 0.1$ \\
\cmidrule(lr){2-6}
\HWC{} & $\Lambda$CDM + running & 7 & 0 & 0 & 0 \\
 & + oscillation & 10 & $-7.4$ & $-1.4$ & $-1.2 \pm 0.1$ \\
\cmidrule[\heavyrulewidth](lr){2-6}
\end{tabular}
\caption{Comparison of the oscillation model with a featureless spectrum, i.e.\ $\Lambda$CDM with a running of the spectral index ($\Aosc = 0$). $N$ is the number of cosmological parameters, the values of $\Delta\chi^2$ are those of the true profile likelihood, and $B$ is the Bayes factor for the prior $0 \leq \Aosc \leq 0.1$.}
\label{tab:model_comparison}
\renewcommand{\arraystretch}{1.0}
\end{table}

The Bayes factor leads to the same conclusion. The evidence of the oscillation model, computed from the combined emulator, is $\ln Z = -6234.97 \pm 0.11$, and that of $\Lambda$CDM with running is $\ln Z = -6233.98 \pm 0.10$, so that $\ln B = -1.0 \pm 0.1$. On the Jeffreys scale this lies at the border between inconclusive and weak evidence against the oscillation \cite{Trotta:2008qt}, although the best fit lowers $\chi^2$ by 10.1. This is because the improvement is confined to a small part of the prior volume of the oscillation parameters. The Savage--Dickey density ratio gives a similar value, and the dependence of $B$ on the prior on $\Aosc$ is discussed in Appendix~\ref{app:evidence}.

\subsection{Adding small-scale CMB data: ACT DR6 and SPT-3G D1 (\texorpdfstring{\HWC{}}{H0WC})}
\label{sec:actspt}

The ground-based measurements of the CMB temperature and polarisation anisotropies by ACT \cite{AtacamaCosmologyTelescope:2025blo} and SPT \cite{SPT-3G:2025bzu} now complement Planck in the damping tail. An oscillation in $\ln k$ appears at all multipoles, and the small-scale spectra test the features found with Planck with independent data. We use the baseline data combination of the $H_0$ World Cup (\HWC{}) \cite{Schoneberg:2026eys,Schoneberg:2026vaf}, again through \cobaya{}:
\begin{itemize}
    \item \textbf{Planck}: the low-$\ell$ temperature likelihood (Commander) as before, and the low-$\ell$ $E$-mode polarisation likelihood built on the SRoll2 maps \cite{Delouis:2019bub,Pagano:2019tci} in place of SimAll. The high-$\ell$ CamSpec PR4 likelihood \cite{Efstathiou:2019mdh,Rosenberg:2022sdy} is restricted to $\ell < 1000$ in TT and $\ell < 600$ in TE and EE, i.e.\ the small scales are left to the ground-based data.
    \item \textbf{ACT DR6}: the foreground-marginalised likelihood of the CMB-only TT, TE and EE bandpowers, ACT-lite \cite{AtacamaCosmologyTelescope:2025blo}, at $600 \leq \ell \leq 6500$.
    \item \textbf{SPT-3G D1}: the TT, TE and EE likelihood of the 2019 and 2020 observations of the SPT-3G Main field \cite{SPT-3G:2025bzu}, in its foreground-marginalised form \cite{Balkenhol:2024obe}, as implemented in candl \cite{Balkenhol:2024sbv}.
    \item \textbf{CMB lensing}: the ACT DR6 lensing power spectrum combined with that of Planck PR4 \cite{ACT:2023dou,ACT:2023kun,Carron:2022eyg}, and the SPT-3G lensing power spectrum of the 2019 and 2020 polarisation data, reconstructed with the MUSE method \cite{SPT-3G:2024atg}.
    \item \textbf{DESI BAO} and \textbf{Pantheon+}, as before \cite{DESI:2025zgx,Scolnic:2021amr,Brout:2022vxf}.
\end{itemize}
The nine nuisance parameters of CamSpec are sampled as before, and the overall calibration of ACT is set equal to that of Planck. The polarisation efficiency of ACT has a uniform prior, and the temperature and polarisation calibrations of SPT have the priors of the combination of the SPT collaboration with Planck and ACT \cite{SPT-3G:2025bzu}. This gives 22 sampled parameters.

\subsubsection{Numerical setup}
\label{sec:actspt:setup}

The ACT-lite likelihood requests the lensed spectra up to $\ell = 9000$. However, its $\chi^2$ does not change when the spectra are computed only up to $\ell = 6327$, which means that it does not use the higher multipoles, and we therefore compute the spectra only up to $\ell = 6327$. We use the precision settings that \CLASSpp{} provides for ACT DR6, with one exception. They solve the perturbations up to $k = 10\,\mathrm{Mpc}^{-1}$ for the non-linear power spectrum, and we keep the default of $2\,\mathrm{Mpc}^{-1}$. In $\ell$ and $k$ we use the denser sampling of Sec.~\ref{sec:numerics:classpp} throughout, and at the maxima of the likelihood below $\omega = 200$ the improvement in $\chi^2$ then differs from that with a converged sampling by at most 0.18 (Appendix~\ref{app:classpp}). The accurate computation of the lensed spectra is essential here. With the fast one the imprint of an oscillation at $\omega = 50$ on spectra with the noise of ACT is wrong by roughly half. One evaluation of the likelihood takes about 78 core seconds, compared with 24 for \PDP{}.

The bandpowers of ACT and SPT are binned in $\ell$, with $\Delta\ell = 50$ at $\ell \lesssim 2000$ for ACT and up to $\ell = 4000$ for SPT, and with wider bins at higher multipoles for ACT. Since the period of the oscillation in $\ell$ is $2\pi\ell/\omega$, the binning removes most of the imprint of a fast oscillation. At $\omega = 100$ the $\chi^2$ of the imprint in the bandpowers of ACT and SPT is about 29\% and 23\% of what the same data would give multipole by multipole, and at $\omega = 200$ about 7\% and 3\% (Appendix~\ref{app:classpp}). This does not bias the likelihood, which compares the theory with the bandpowers through their window functions. A finer binning is not an option, since the bandpowers of ACT and SPT, their covariances and the treatment of filtering and mode coupling are validated at this binning \cite{AtacamaCosmologyTelescope:2025blo,SPT-3G:2025bzu}. However, at high frequency the sensitivity of \HWC{} comes mainly from the Planck spectra at $\ell < 1000$, and at $\omega = 200$ it is about half that of \PDP{}, whose CamSpec likelihood is not binned. At $\omega = 60$, on the other hand, \HWC{} is the more sensitive of the two.

The emulators of \HWC{} are set up and trained as those of \PDP{}. The prior volume is split into eight parts instead of seven, and the few differences in the split and in the settings are given in Appendices~\ref{app:split} and \ref{app:boosts}. In total the training required \ensuremath{1.1\times10^{6}} evaluations of the true likelihood. The combined emulator reproduces the true log-likelihood at 500 samples of its posterior with an rms error of 0.16 (0.07 for \PDP{}, Appendix~\ref{app:accuracy}).

\subsubsection{Parameter inference}
\label{sec:actspt:params}

\begin{figure}[tb]
\centering
\includegraphics[width=\textwidth]{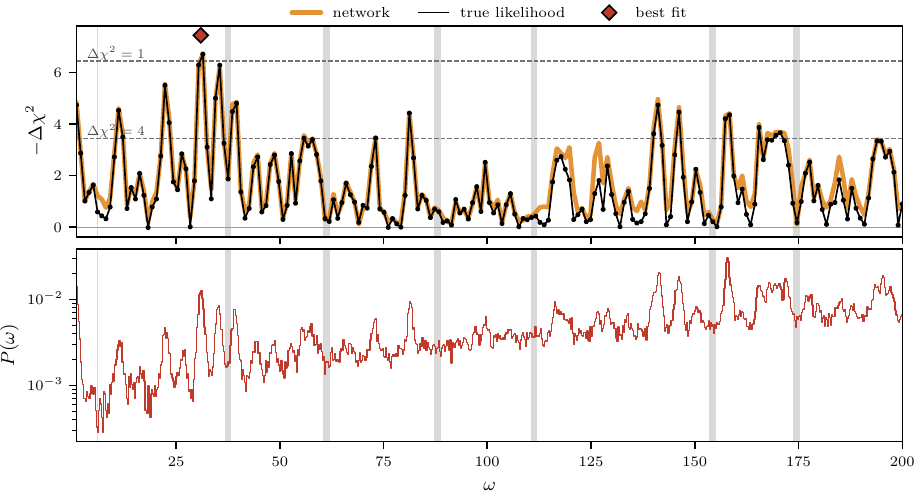}
\caption{The frequency of the oscillation for \HWC{}, as in Fig.~\ref{fig:omega}: the profile likelihood calculated with the true likelihood (black) and with the emulator (orange), and the marginalised posterior. As for \PDP{}, the best fit (diamond) lies between two nodes of the grid in $\omega$, and the profile on the grid therefore lies slightly below it. Neighbouring emulators of \HWC{} are blended in the grey bands.}
\label{fig:omega_wc}
\end{figure}

The small-scale CMB data does not support the feature found with \PDP{}, i.e.\ the oscillation at $\omega \approx 51$ of its best fit (Fig.~\ref{fig:omega_wc}). The peak at this frequency, which contains 12\% of the posterior with \PDP{}, contains only 0.6\% with \HWC{}, less than an even share of 1.8\%. This means that the data of \HWC{} has no preference for this frequency. The feature may be due to the Planck spectra at the multipoles which \HWC{} takes from ACT and SPT, or it may be present in Planck at lower multipoles and be counteracted by ACT and SPT. With CamSpec PR4 the improvement at this frequency accumulates mainly at $\ell \approx 800$--1500 \cite{Raffaelli:2026bhk}, i.e.\ largely at multipoles which \HWC{} takes from ACT and SPT.

\begin{figure}[tb]
\centering
\includegraphics[width=\textwidth]{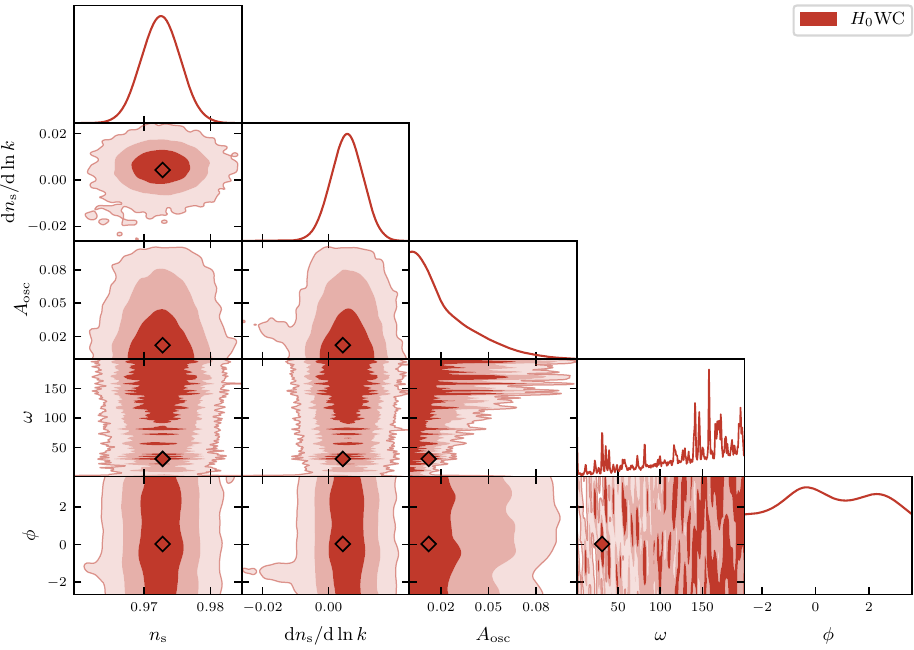}
\caption{The posterior of the spectral index, the running and the oscillation parameters for \HWC{} (68\%, 95\% and 99.7\% regions), as in Fig.~\ref{fig:triangle}. The panels with $\omega$ are drawn from histograms in bins of $\Delta\omega = 0.5$.}
\label{fig:triangle_wc}
\end{figure}

With \HWC{} the posterior is spread over even more peaks than with \PDP{}, and 86\% of it lies above $\omega = 60$ (Fig.~\ref{fig:omega_wc}). The peak at $\omega \approx 196$, the second highest maximum of \PDP{}, is also present. However, the two datasets are not independent, since they share the Planck spectra at $\ell < 600$--1000.

The amplitude is limited to $\Aosc < 0.062$ at 95\%~CL (Table~\ref{tab:params} and Fig.~\ref{fig:triangle_wc}), close to the limit for \PDP{}, since again the high frequencies allow larger amplitudes, and there the binning of ACT and SPT reduces their sensitivity further (Sec.~\ref{sec:actspt:setup}). For $\omega \leq 60$, on the other hand, the amplitude is more strongly constrained than with \PDP{}, and we find $\Aosc < 0.019$ (0.031 for \PDP{}). The other constraints are listed in Table~\ref{tab:params}. As with \PDP{} the posteriors of the $\Lambda$CDM parameters are close to Gaussian and the running is consistent with zero, but $n_\mathrm{s}$ is larger, as expected for data that include ACT \cite{AtacamaCosmologyTelescope:2025blo}.

\needspace{6\baselineskip}
\subsubsection{Profile likelihoods and model comparison}
\label{sec:actspt:profiles}

\begin{figure}[tb]
\centering
\includegraphics[width=\textwidth]{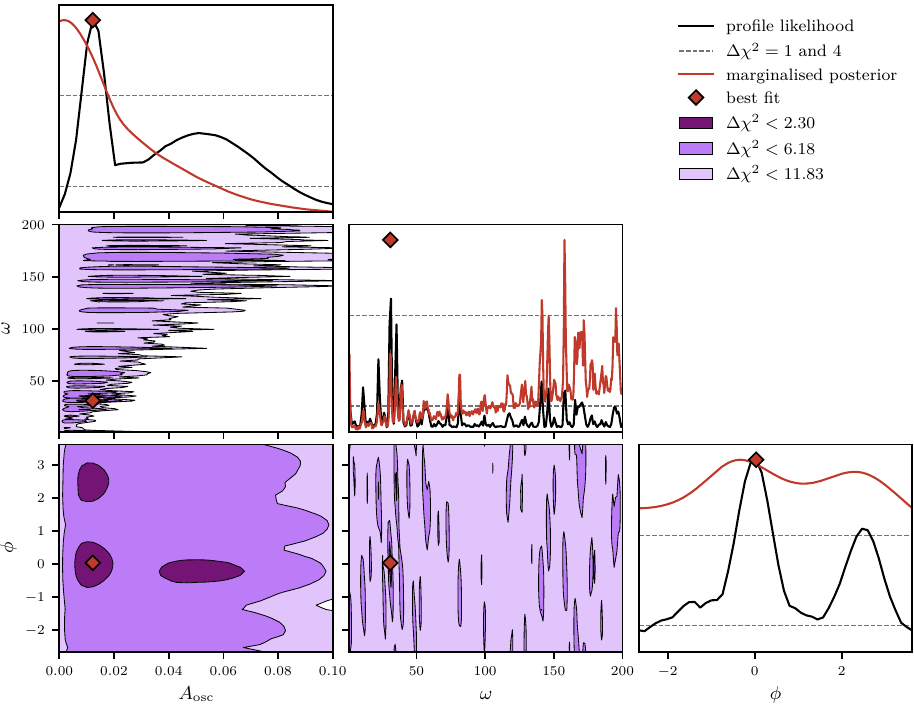}
\caption{The profile likelihood of \HWC{} in the oscillation parameters, as in Fig.~\ref{fig:profile2d}. In $\omega$ the best fit again lies between two nodes of the grid, so that the one-dimensional profile is below 1. In the two lowest windows of $(\Aosc, \omega)$ the grid ends at $\Aosc = 0.04$ and 0.07, respectively (white above). The marginalised posterior on the diagonal is shown in red.}
\label{fig:profile2d_wc}
\end{figure}

The profile likelihood confirms that the feature at $\omega \approx 51$ is not supported (Figs.~\ref{fig:omega_wc} and \ref{fig:profile2d_wc}). At this frequency the improvement in the profile of \HWC{} is at most 1.8, compared with 10.1 for \PDP{}. Furthermore, with the amplitude of the best fit of \PDP{}, $\chi^2$ at this frequency is 7.2 larger than without oscillation for any phase. This means that the feature found with \PDP{} is disfavoured by the data of \HWC{}.

The best fit of \HWC{} lies at $\omega = 31.0$ and $\Aosc = 0.012$, and it lowers $\chi^2$ by 7.4 compared with the best fit without oscillation, i.e.\ $\Delta\mathrm{AIC} = -1.4$ (Table~\ref{tab:model_comparison}). It is only slightly better than the next maxima, at $\omega \approx 36$ and 22 (Fig.~\ref{fig:omega_wc}). Since the improvement is smaller than with \PDP{} and the number of maxima is similar, the look-elsewhere effect leaves it even less significant, and 3.4 maxima as high as the best fit are expected without any oscillation.

For the amplitude the profile and the posterior differ as for \PDP{}. The profile disfavours $\Aosc = 0$ by $\Delta\chi^2 = 7.4$, but the posterior is largest there. At large amplitude the profile stays within $\Delta\chi^2 = 4$ of the best fit up to $\Aosc = 0.084$, and the posterior limit is $\Aosc < 0.062$. The Bayes factor of \HWC{}, computed as for \PDP{}, is $\ln B = -1.2 \pm 0.1$, i.e.\ weak evidence against the oscillation (Appendix~\ref{app:evidence}).

\section{Discussion and conclusions}
\label{sec:conclusions}

We have analysed logarithmic oscillations in the primordial power spectrum, as predicted by axion monodromy inflation, together with a running of the spectral index, for frequencies up to $\omega = 200$ and with all cosmological, oscillation and nuisance parameters free. We have used two combinations of current data: a conservative one of the full Planck temperature, polarisation and PR4 lensing data together with DESI DR2 BAO and Pantheon+ supernovae, and a more aggressive one in which the small-scale CMB spectra come from ACT and SPT.

Our main conclusion is that current data shows no robust oscillatory feature. With the conservative data the profile likelihood has a fairly well-defined best fit at a frequency $\omega \simeq 51$, which lowers $\chi^2$ by 10.1 with respect to a featureless spectrum. However, the more aggressive data does not support it. At this frequency its profile likelihood improves $\chi^2$ by at most 1.8, and with the amplitude of the conservative best fit $\chi^2$ is higher than for a featureless spectrum. This may be due to the Planck spectra at the multipoles which the more aggressive data takes from ACT and SPT, or the feature may be present in Planck at lower multipoles and be counteracted by ACT and SPT. The two datasets share the Planck spectra at large scales, and the difference between them lies mainly at small scales.

The model comparison leads to the same conclusion. In terms of the AIC the improvement is a mild preference, and the Bayes factors, $\ln B = -1.0$ and $-1.2$, weakly favour the featureless spectrum for both datasets. Furthermore, the conservative data has maxima of almost the same height at high frequency, and once this look-elsewhere effect is taken into account a maximum as high as the best fit is expected even without any oscillation. Improvements of this size are also known to occur in fits to featureless simulated data. The amplitude is limited to $\Aosc < 0.056$ with the conservative data and $\Aosc < 0.062$ with the more aggressive data (95\%~CL), mainly by the high frequencies.

Our results agree well with the recent analysis of the Planck spectra by Raffaelli et al.~\cite{Raffaelli:2026bhk}. With the CamSpec PR4 likelihood they find the best fit of the logarithmic template at the same frequency, $\omega \simeq 51$, with the same amplitude and the same improvement in $\chi^2$, and they also find that it is not significant, both after the look-elsewhere correction and in terms of the Bayes factor. This means that neither the DESI BAO and Pantheon+ data nor the running of the spectral index, which their analysis does not include, changes the feature. Their analysis uses the Planck data alone and a smaller range of frequencies, with a prior uniform in $\log\omega$. Our analysis extends this to frequencies up to $\omega = 200$, where the Planck data has maxima of almost the same height, and to the small-scale data of ACT and SPT, which does not support the feature.

Our second conclusion concerns the method. With \client{} the posteriors, the Bayesian evidence and the profile likelihoods calculated with the true likelihood are all available for a likelihood with many narrow maxima in frequency, and we have been able to run all parts of the analysis of this very complex model with high numerical precision. Such a full statistical analysis is not feasible by calling the Einstein--Boltzmann code directly (Appendix~\ref{app:cost}). The comparison of the profile likelihood and the posterior of the amplitude shows why both are needed. The profile disfavours $\Aosc = 0$, but the posterior is largest there, since at most frequencies and phases the data allows only a small amplitude. At large amplitude, on the other hand, the profile allows larger amplitudes than the posterior, since at high frequency the imprint of the oscillation on the spectra is small, but only near narrow maxima with a very small volume. Both differences are prior volume effects.

In order to make this analysis possible the prior volume was split in $\omega$ between several emulators, which were combined into a single one, and the profile likelihoods were refined with the true likelihood. The combined emulators reproduce the true log-likelihood at samples of their posteriors with rms errors of 0.07 and 0.16, so that the posteriors need no correction. In conclusion, current data shows no robust evidence for logarithmic oscillations in the primordial spectrum, and emulating the likelihood itself makes the full statistical analysis of such features affordable. \client{} is publicly available at \clienturl{}. It would be very interesting to extend it to a drifting frequency and to the joint analysis of the power spectrum and the bispectrum, eq.~\eqref{eq:fnl}.

\section*{Reproducibility}

The analysis uses \client{} \cite{Janken:2025wlq}, the nested sampler BEST \cite{Nygaard:2026fgl,Nygaard:2026hbr}, \cobaya{} \cite{Torrado:2020dgo} and \CLASSpp{} with an external primordial spectrum. The emulators were trained with the scripts for cluster runs that come with \client{}, and they were combined, integrated and profiled with its tools \texttt{combine}, \texttt{evidence} and \texttt{profiles}, with the settings of Tables~\ref{tab:client} and \ref{tab:client_b}. The run settings, the \cobaya{} input files and the scripts which produce the tables and figures in this paper are publicly available, together with a description of how to reproduce the results, in the branch \texttt{papers/2026-axion-monodromy} of the \client{} repository: \url{https://github.com/AarhusCosmology/client-f/tree/papers/2026-axion-monodromy}.

\section*{Acknowledgements}
We thank Luca Janken for important contributions to \client{} and valuable feedback on the paper. We thank Mario Ballardini for valuable comments on the paper. We acknowledge computing resources from the Centre for Scientific Computing Aarhus (CSCAA).

\appendix

\section{The \client{} emulators}
\label{app:client}

This appendix describes the steps of \client{} used in this work, in the order in which they are carried out. As in the main text, the errors of the network and the depths below the best point that steer its training are given in $\ln\mathcal{P}$, i.e.\ half of their values in $\chi^2$, and the values of the profile likelihoods in $\chi^2$. The settings are listed in Tables~\ref{tab:client} and \ref{tab:client_b}. They are the same for all emulators of both datasets, which differ only in their likelihoods and training boxes, apart from the smaller share of the boost points given to light modes for \HWC{} (Appendix~\ref{app:boosts}).

\begin{table}[tb]
\centering
\small
\setlength{\tabcolsep}{5pt}
\begin{tabular}{llp{5.6cm}l}
\cmidrule[\heavyrulewidth](lr){2-4}
 & Setting & Meaning & Value \\
\cmidrule(lr){2-4}
\texttt{prior} & \texttt{n\_samples} & points of the Latin hypercube design & 3\,000 \\
 & \texttt{n\_inject} & points from a Gaussian fit to it & 100 \\
\cmidrule(lr){2-4}
\texttt{acquisition} & \texttt{n\_append} & new points per iteration & 3\,000 \\
 & \texttt{target\_temperature} & temperature $T$ of the chain & 7 \\
 & \texttt{pool\_factor} & candidates per selected point & 20 \\
 & \texttt{n\_neighbors} & neighbours in the density estimates & 20 \\
 & \texttt{n\_inject} & Gaussian points per iteration & 100 \\
 & \texttt{inject\_window} & fitted points, in units of $Tn/2$ & 16 \\
\cmidrule(lr){2-4}
\texttt{acquisition.audit} & \texttt{tau} & check maxima this far above the best & 5 \\
 & \texttt{chain} & boosts: also the cooled chain & yes \\
 & \texttt{posterior} & boosts: posterior samples checked & 150 \\
\cmidrule(lr){2-4}
\texttt{convergence} & \texttt{threshold} & largest scatter, in units of $T$ & 1.2 \\
 & \texttt{consecutive} & for this many networks in a row & 2 \\
\cmidrule(lr){2-4}
\texttt{acquisition.boost} & \texttt{boosts} & number of boosts & 10 \\
 & \texttt{n} & posterior points per boost & 10\,000 \\
 & \texttt{density\_filter} & select where the training set is sparse & yes \\
 & \texttt{mixing} & merge basins below this memory & 0.1 \\
 & \texttt{min\_share} & smallest share of walkers in a mode & 0.005 \\
 & \texttt{balance\_floor} & modes below this get fewer points & 0.1 (\HWC{}) \\
 & \texttt{mode\_depth} & drop modes this far below the highest & $Tn/2$ \\
 & \texttt{background} & function fitted near the maximum & quadratic \\
 & \texttt{train\_domain} & training domain, in units of its radius & 1.25 \\
 & \texttt{region} & region edge, in units of the reach & 1.5 \\
 & \texttt{region\_quantile} & quantile that sets the reach & 0.999 \\
\cmidrule[\heavyrulewidth](lr){2-4}
\end{tabular}
\caption{The settings of \client{} that determine the training points, by their names in the configuration files of the code. $T$ is the temperature of the tempered posterior and $n$ the number of sampled parameters. The domain, the region and its reach are defined in Appendix~\ref{app:boosts}.}
\label{tab:client}
\end{table}

\begin{table}[tb]
\centering
\small
\setlength{\tabcolsep}{5pt}
\begin{tabular}{llp{5.6cm}l}
\cmidrule[\heavyrulewidth](lr){2-4}
 & Setting & Meaning & Value \\
\cmidrule(lr){2-4}
\texttt{model} & \texttt{n\_layers} & hidden layers & 5 \\
 & \texttt{n\_neurons} & neurons per hidden layer & 512 \\
 & \texttt{activation} & activation function of ref.~\cite{SPECULATOR:2020} & alsing \\
\cmidrule(lr){2-4}
\texttt{training} & \texttt{loss} & error relative to the depth & msre \\
 & \texttt{sigma\_level} & level (in $\sigma$) that sets the margin $c$ & 3 \\
 & \texttt{msre\_ess\_floor} & smallest effective fraction of points & 0.1 \\
 & \texttt{learning\_rate} & initial learning rate (Adam) & $10^{-4}$ \\
 & \texttt{lr\_schedule} & $\times 0.3$ if no improvement in 75 epochs & plateau \\
 & \texttt{batch\_size} & points per batch & 128 \\
 & \texttt{warm\_start} & the first network starts from the previous one & yes \\
 & \texttt{members} & trainings per network in the boosts & 12 \\
 & \texttt{member\_pilot} & posterior points of each network checked & 100 \\
 & \texttt{member\_margin} & lower error needed to replace the first & 0.2 \\
\cmidrule(lr){2-4}
\texttt{sampling} & \texttt{n\_walkers} & walkers of the ensemble sampler & 432 \\
 & \texttt{n\_steps} & steps, after a burn-in of 5\,000 & 60\,000 \\
\cmidrule(lr){2-4}
\texttt{profiles} & \texttt{profile\_points} & nodes per parameter & 50 \\
 & \texttt{surrogate\_optimizer} & scan; legacy for $(\Aosc, \phi)$ & hybrid \\
 & \texttt{hybrid\_hops} & most rounds of the hybrid scan & 20 \\
 & \texttt{corrected\_method} & refinement with the true likelihood & newton \\
 & \texttt{maxfun} & most true evaluations per node & 150 \\
 & \texttt{region} & the emulator inside its region & run \\
\cmidrule(lr){2-4}
\texttt{evidence} & \texttt{-{}-n-live} & live points of BEST & 2048 \\
 & \texttt{-{}-n-live-updates} & live points replaced per iteration & 128 \\
 & \texttt{-{}-tolerance} & evidence left in the live points & $10^{-3}$ \\
\cmidrule[\heavyrulewidth](lr){2-4}
\end{tabular}
\caption{The settings of the networks and their training, of the tempered chains, of the profile likelihoods and of the evidence (the options of the evidence tool of \client{}).}
\label{tab:client_b}
\end{table}

\subsection{The iterative stage}
\label{app:training}

In \client{} a neural network learns the log-posterior returned by \cobaya{}, the sum of the log-likelihoods and the log-prior, as a function of all sampled parameters. The emulator is a fully connected network with five hidden layers of 512 neurons and the trainable activation function of ref.~\cite{SPECULATOR:2020}, $f(x) = [\gamma + (1 - \gamma)\,\sigma(\beta x)]\,x$, where $\sigma(y) = (1 + \e^{-y})^{-1}$ is the logistic function and $\beta$ and $\gamma$ are trained for each neuron. The inputs of the network are standardised, its output is scaled with the mean and the standard deviation of the training values, and outside the training box the emulated posterior is zero. The loss function is the mean squared error relative to the depth below the best training point, $\ln\mathcal{P}_\mathrm{max} - \ln\mathcal{P} + c$, so that the network is most accurate near the maximum. The margin $c$ is the depth in $\ln\mathcal{P}$ within which a Gaussian posterior contains 99.7\% of its probability, i.e.\ half the $3\sigma$ quantile of a $\chi^2$ distribution with $n$ degrees of freedom, where $n$ is the number of parameters. Early in the training, when few points lie near the maximum, $c$ is increased until the effective number of training points in the loss is at least a tenth of all points. A new network starts from the weights of the previous one, adjusted to the new scaling of the inputs and the output so that it reproduces the previous network before the training begins. The learning rate is reduced when the loss at the validation points, a tenth of the training set, stops decreasing.

The training starts from 3000 points of a Latin hypercube in the training box and 100 points drawn from a Gaussian fitted to them. The network of an iteration is sampled with an ensemble sampler at the temperature $T = 7$, and 3000 new points are selected from the tempered chain. At a pool of candidates from the chain, the density of the tempered posterior and that of the training set are both estimated from the distances to the 20 nearest neighbours, and the candidates are selected with a probability proportional to the deficit of the training set. The new points therefore go mainly where the training set is sparser than the tempered posterior. A further 100 points are drawn from a Gaussian fitted to the training points within $16\,Tn/2$ of the best point in $\ln\mathcal{P}$. Points at which \CLASSpp{} fails are given a fixed low value, the median of the values of the initial design, so that the network learns that these regions are strongly disfavoured, and they are not used in the estimates of the error of the network.

Before each tempered chain is run, the network is audited. Its maxima are found by gradient ascent from several thousand starting points, and those which the network places more than 5 above the best point in $\ln\mathcal{P}$ are evaluated with the true likelihood, at most 60 per round in at most five rounds. After a round the network is trained briefly on the new points, so that a maximum which is not present in the true likelihood is removed. In the boosts the audit also follows walkers of the cooled tempered chain to their maxima, and it compares the network with the true likelihood at 150 samples of its posterior. If the network is too high there on average, by more than 2 in $\ln\mathcal{P}$, it is trained further on these points.

The first stage ends when the standard deviation of the difference between the network and the true likelihood at its new tempered points, which the network selected from its own chain and has not been trained on, is below $1.2\,T = 8.4$ in $\ln\mathcal{P}$ for two consecutive networks. For the emulators of \PDP{} this happened after network 3 to 5, and for those of \HWC{} after network 3 to 5 (Tables~\ref{tab:training_pp} and \ref{tab:training_wc}).

\subsection{Boosts}
\label{app:boosts}

In the boosts the points are selected from the posterior at $T = 1$, and its modes are identified first. 512 walkers from the tempered chain are cooled to $T = 1$ and climb to the nearest maxima of the network, maxima connected by a path along which the network falls by less than 1 in $\ln\mathcal{P}$ form a basin, and two basins are one mode if a chain at $T = 1$ moves freely between them. Modes lower than the highest by more than $Tn/2$ are dropped, and the evidence of each mode is estimated by bridge sampling (Sec.~\ref{sec:numerics:inference}). All emulators found a single mode in all boosts, except the emulator of \HWC{} at low frequency, which found 1 to 3 (two ellipsoids of its region at the end).

All modes have the same weight, and the 10\,000 points of a boost are selected where the training set falls short of this distribution. For \HWC{} a mode with a weight $w < 0.1$ receives only the fraction $w/0.1$ of an equal share. From the training after the first boost on, the network learns the deviation from a quadratic function fitted to the training points within $Tn$ of the best point, with $T = 7$, and it is trained only on points inside a sphere, after whitening with their covariance, of radius 1.25 times their largest distance from their mean. Otherwise values many thousands below the maximum would set the scale of its output.

From the second boost on, the emulator is restricted to a region of ellipsoids around the boost points within $Tn/2$ of the best point, each with the mean and the covariance of its points, and a point lies inside an ellipsoid within 1.5 times the 99.9\% quantile of the Mahalanobis distances of these points. Outside the region it is replaced by a value 1000 below the best point, which decreases with the distance. The tempered chain, the search for modes, the audit, the evidence and the profiles all see the emulator through this region.

For \PDP{} the posterior of each emulator after four boosts, when the emulators had been trained on 430\,559 points, agrees with that of its final network: the means differ by at most 0.06 standard deviations, the widths by at most 5\% and the evidences by at most 0.03 in $\ln Z$. However, the error at samples of the posterior continued to decrease, from between 0.07 and 0.24 in $\ln\mathcal{P}$ to between 0.04 and 0.09 after ten boosts (Table~\ref{tab:training_pp}). For \HWC{} the same holds after five boosts and 590\,891 points (at most 0.15 standard deviations, 9\% and 0.04 in $\ln Z$), with errors decreasing from between 0.11 and 0.38 to between 0.07 and 0.20 (Table~\ref{tab:training_wc}). In the boosts a network keeps its weights at the end of the training, since a network that starts from the previous one has already been trained on most of its validation points.

In the boosts each network is trained 12 times, once starting from the previous network and otherwise from random weights. Depending on its start, a training either resolves the fine structure of the likelihood or smooths it out, and a network that starts from the previous one can keep a maximum that the next boost does not reveal. Each network therefore draws 100 points from its own posterior, and the true likelihood is computed at all 1\,200 points. The error of a network is its rms difference from the true likelihood at the points of this pool that it or the true likelihood places within $c$ of the best point, where $c$ is the margin of the loss function (Appendix~\ref{app:training}). The network that starts from the previous one is kept unless another has an error lower by more than 20\% (at the first training after the first boost, the one with the lowest error), and the points of the pool are added to the training set.

In a first emulator of \PDP{} for $2.5 \leq \omega \leq 200$, before this range was split (Table~\ref{tab:zones}), the network that started from the previous one was rejected in six consecutive boosts, since it was too high by 27 to 53 in $\ln\mathcal{P}$ at its own posterior samples, while the kept networks from random weights were within 0.2. In the zones it was usually kept once the emulator had found its maxima. Over all emulators of both datasets, including those that were split later, the error of the kept network at the points of its own boost lies within 30\% of its error at the pilot points for 87\% of the 170 networks, and exceeds it by more than 30\% for only 4\%.

\begin{table}[tb]
\centering
\renewcommand{\arraystretch}{1.1}
\begin{tabular}{rlccccccc}
\cmidrule[\heavyrulewidth](lr){2-9}
 & & \multicolumn{7}{c}{\textbf{Part} (Table~\ref{tab:zones})} \\
 & & low & 1a & 1b & 2a & 2b & 3 & 4 \\
\cmidrule(lr){2-9}
Iterations & network 0 & 72 & 69 & 44.7 & 168 & 50.2 & 113 & 51.3 \\
 & network 1 & 18.3 & 21.4 & 12.7 & 23.7 & 12.8 & 16.2 & 21.2 \\
 & network 2 & 13.7 & 13 & 8.07 & 9.78 & 7.69 & 6.39 & 8.07 \\
 & network 3 & 8.38 & 9.63 & 5.53 & 7.88 & 7.59 & 5.43 & 5.64 \\
 & network 4 & 7.53 & 5.99 & & 4.58 & & & \\
 & network 5 & & 6.03 & & & & & \\
\cmidrule(lr){2-9}
Boosts & 1 & 2.4 & 1.4 & 2 & 1 & 1.4 & 1.6 & 1.5 \\
 & 2 & 0.44 & 0.82 & 0.33 & 0.81 & 1.2 & 1 & 1.3 \\
 & 3 & 0.28 & 0.5 & 0.15 & 0.27 & 0.26 & 0.87 & 1.3 \\
 & 4 & 0.23 & 0.28 & 0.12 & 0.16 & 0.12 & 0.48 & 0.47 \\
 & 5 & 0.15 & 0.22 & 0.093 & 0.089 & 0.084 & 0.24 & 0.19 \\
 & 6 & 0.15 & 0.2 & 0.066 & 0.068 & 0.072 & 0.14 & 0.14 \\
 & 7 & 0.1 & 0.17 & 0.057 & 0.061 & 0.063 & 0.11 & 0.13 \\
 & 8 & 0.07 & 0.14 & 0.049 & 0.056 & 0.056 & 0.091 & 0.12 \\
 & 9 & 0.087 & 0.27 & 0.051 & 0.051 & 0.049 & 0.087 & 0.11 \\
 & 10 & 0.06 & 0.12 & 0.042 & 0.048 & 0.048 & 0.079 & 0.11 \\
\cmidrule(lr){2-9}
Posterior & 1 & 4.2 & 1.4 & 2 & 1 & 1.2 & 2.5 & 2.8 \\
 & 2 & 0.37 & 0.79 & 0.32 & 0.89 & 1.1 & 1 & 1.3 \\
 & 3 & 0.19 & 0.57 & 0.12 & 0.24 & 0.23 & 0.99 & 1.4 \\
 & 4 & 0.19 & 0.26 & 0.096 & 0.16 & 0.1 & 0.43 & 0.34 \\
 & 5 & 0.099 & 0.24 & 0.07 & 0.1 & 0.07 & 0.2 & 0.17 \\
 & 6 & 0.096 & 0.19 & 0.095 & 0.065 & 0.063 & 0.12 & 0.15 \\
 & 7 & 0.061 & 0.21 & 0.058 & 0.048 & 0.057 & 0.088 & 0.12 \\
 & 8 & 0.058 & 0.13 & 0.047 & 0.053 & 0.053 & 0.086 & 0.13 \\
 & 9 & 0.048 & 0.14 & 0.041 & 0.05 & 0.049 & 0.079 & 0.11 \\
 & 10 & 0.041 & 0.092 & 0.039 & 0.06 & 0.042 & 0.07 & 0.11 \\
 & final & 0.077 & 0.084 & 0.039 & 0.049 & 0.043 & 0.072 & 0.087 \\
\cmidrule[\heavyrulewidth](lr){2-9}
\end{tabular}
\caption{The difference between the true and the emulated log-likelihood for the parts of \PDP{}, at points on which each network had not been trained. The first two blocks give its standard deviation at the tempered points selected by each network of the iterative stage and at the posterior points of each boost. The last block gives the rms difference, which also includes the mean of the difference, at 500 samples of the posterior of the network that selected the points of each boost, and of the final network, trained on the points of all ten boosts. In the boosts each network is the one kept of its 12 trainings (Appendix~\ref{app:boosts}). }
\label{tab:training_pp}
\renewcommand{\arraystretch}{1.0}
\end{table}

\begin{table}[tb]
\centering
\renewcommand{\arraystretch}{1.1}
\begin{tabular}{rlcccccccc}
\cmidrule[\heavyrulewidth](lr){2-10}
 & & \multicolumn{8}{c}{\textbf{Part} (Table~\ref{tab:zones})} \\
 & & low & 1a & 1b & 2a & 2b & 3 & 4a & 4b \\
\cmidrule(lr){2-10}
Iterations & network 0 & 48.5 & 33.1 & 32.1 & 40 & 37.7 & 26.9 & 107 & 36.1 \\
 & network 1 & 24.2 & 13.2 & 12.6 & 11.1 & 11.4 & 12.6 & 13.2 & 11.1 \\
 & network 2 & 14.3 & 9.92 & 8.93 & 10.1 & 5.89 & 6.5 & 12.1 & 7.52 \\
 & network 3 & 9.81 & 8.86 & 7.12 & 6.37 & 4.73 & 4.72 & 6.8 & 5.3 \\
 & network 4 & 7.06 & 7.54 & 5.31 & 5.59 & & & 6.07 & \\
 & network 5 & 6.47 & 6.27 & & & & & & \\
\cmidrule(lr){2-10}
Boosts & 1 & 2.1 & 2.1 & 1.6 & 1.8 & 1.7 & 2.2 & 2.5 & 1.4 \\
 & 2 & 0.49 & 1.6 & 1.1 & 1.1 & 0.69 & 1.1 & 1.1 & 0.7 \\
 & 3 & 0.35 & 1.7 & 0.39 & 0.88 & 0.66 & 1 & 0.45 & 0.56 \\
 & 4 & 0.24 & 0.45 & 0.23 & 0.28 & 0.55 & 1 & 0.38 & 0.55 \\
 & 5 & 0.25 & 0.27 & 0.16 & 0.19 & 0.26 & 0.97 & 0.36 & 0.38 \\
 & 6 & 0.19 & 0.18 & 0.12 & 0.15 & 0.16 & 0.38 & 0.3 & 0.34 \\
 & 7 & 0.18 & 0.14 & 0.12 & 0.14 & 0.13 & 0.3 & 0.25 & 0.28 \\
 & 8 & 0.11 & 0.12 & 0.092 & 0.12 & 0.13 & 0.28 & 0.18 & 0.26 \\
 & 9 & 0.12 & 0.12 & 0.085 & 0.094 & 0.13 & 0.25 & 0.15 & 0.26 \\
 & 10 & 0.094 & 0.1 & 0.075 & 0.085 & 0.12 & 0.25 & 0.15 & 0.22 \\
\cmidrule(lr){2-10}
Posterior & 1 & 3.4 & 2.8 & 1.7 & 1.7 & 1.9 & 2.5 & 3.6 & 1.5 \\
 & 2 & 0.87 & 1.6 & 1.2 & 1.1 & 0.71 & 1.1 & 1.1 & 0.62 \\
 & 3 & 0.28 & 1.9 & 0.37 & 0.75 & 0.65 & 1.3 & 0.47 & 0.53 \\
 & 4 & 0.25 & 0.42 & 0.19 & 0.25 & 0.6 & 1 & 0.39 & 0.47 \\
 & 5 & 0.22 & 0.22 & 0.18 & 0.18 & 0.22 & 0.75 & 0.34 & 0.37 \\
 & 6 & 0.26 & 0.17 & 0.11 & 0.15 & 0.16 & 0.33 & 0.32 & 0.38 \\
 & 7 & 0.11 & 0.19 & 0.11 & 0.14 & 0.14 & 0.3 & 0.26 & 0.25 \\
 & 8 & 0.11 & 0.15 & 0.083 & 0.1 & 0.12 & 0.29 & 0.2 & 0.24 \\
 & 9 & 0.083 & 0.12 & 0.075 & 0.1 & 0.13 & 0.23 & 0.13 & 0.27 \\
 & 10 & 0.078 & 0.092 & 0.075 & 0.09 & 0.12 & 0.24 & 0.12 & 0.23 \\
 & final & 0.085 & 0.095 & 0.065 & 0.069 & 0.11 & 0.18 & 0.11 & 0.2 \\
\cmidrule[\heavyrulewidth](lr){2-10}
\end{tabular}
\caption{As Table~\ref{tab:training_pp}, for the parts of \HWC{}. }
\label{tab:training_wc}
\renewcommand{\arraystretch}{1.0}
\end{table}

\subsection{Accuracy}
\label{app:accuracy}

In contrast to an emulator of cosmological observables, the accuracy of a likelihood emulator can be measured directly in the region of high posterior probability. We draw 500 samples from the posterior of the combined emulator, compute the true log-likelihood at each of them with \CLASSpp{} and \cobaya{}, and compare the two. For \PDP{} the rms difference is 0.07, and the largest absolute difference is 0.7. In the peak of the best fit, $49 \leq \omega < 52.5$, the rms difference is only 0.02, and above $\omega = 60$, where most of the samples lie, it is 0.07. The median of the samples lies 10.0 below the best training point in $\ln\mathcal{P}$, very close to $n/2$ for $n = 19$ parameters, as expected for samples of a posterior that is close to Gaussian in most directions. Weighting the samples by $\e^{\ln\mathcal{P}_\mathrm{true}-\ln\mathcal{P}_\mathrm{emu}}$ leaves 497 of the 500 samples effective. It moves no posterior mean by more than 0.01 standard deviations and changes no width by more than 1\%, and the correction of the evidence, $\ln\langle\e^{\ln\mathcal{P}_\mathrm{true}-\ln\mathcal{P}_\mathrm{emu}}\rangle$, is 0.002. For \HWC{} the rms difference is 0.16 and the largest absolute difference 1.5. Weighting the samples leaves 487 of the 500 effective, it moves no posterior mean by more than 0.01 standard deviations and changes no width by more than 1\%, and the correction of the evidence is 0.006.

The posterior of a combined emulator is sampled in each cell of its region separately, and the cells are weighted by their evidences, which are estimated by bridge sampling. For \PDP{} \client{} marks 2 of the 7 weights as not reliable, since the estimate has a low efficiency, and for \HWC{} all 9. It then applies no corrections from the true likelihood itself. However, the share of the posterior in the range of each part agrees with that of the nested sampling of the emulator, which integrates the emulator as a whole, to within 0.3 percentage points for \PDP{} and 0.7 for \HWC{}. The corrections above are computed with the weights of the chain.

\subsection{The evidence}
\label{app:evidence}

The evidence is computed from the combined emulator with the nested sampler BEST and the settings of Table~\ref{tab:client_b}. For \PDP{} the error of $\ln Z$ is the standard error of four runs with different seeds, whose scatter is 0.21, and the error that a single run quotes is 0.14. For \HWC{} the evidence of the oscillation model is $\ln Z = -2920.01 \pm 0.03$, and that of $\Lambda$CDM with running is $\ln Z = -2918.79 \pm 0.05$. The featureless model is integrated as the limit of the oscillation model with $\Aosc$ uniform on $[0, 10^{-4}]$, where the likelihood no longer depends on $\omega$ and $\phi$. Restricting $\omega$ to the range of one part of the combined emulator gives evidences which differ by at most 0.2 for \PDP{} and 0.5 for \HWC{}, with standard deviations of 0.08 and 0.18. For comparison, the scatter of single runs of the nested sampler between seeds is 0.19 and 0.10. For \PDP{} the differences between the parts are therefore due to the nested sampling alone. For \HWC{} they are slightly larger and allow an additional error of the emulators of about 0.15 in the evidence of a single part. The errors of $\ln B$ quoted in the text are those of the nested sampling.

The Savage--Dickey density ratio of the posterior at $\Aosc = 0$ gives $\ln B = -1.1$ for \PDP{} and -1.2 for \HWC{}, close to the values from the evidences. The Bayes factor depends on the prior on $\Aosc$, whose range is almost twice the 95\% upper limit, and with $0 \leq \Aosc \leq 0.05$ it would be higher by 0.6 for both datasets.

\subsection{Profile likelihoods}
\label{app:profiles}

The profile tool first recomputes the true likelihood at the three best training points, and it stops if the result differs from the stored values by more than 0.01 in $\chi^2$. In the scan of the emulator, eight walkers per node take random steps, shaped by the curvature of the emulator at its best fit, or jump to the optimum of a neighbouring node. A step is followed by a local maximisation and accepted with the Metropolis rule. The scan stops when two consecutive rounds improve no node, and after at most 20 rounds. The planes $(\Aosc, \omega)$ and $(\omega, \phi)$ are each calculated in four windows in $\omega$ of $50\times50$ nodes, which share their edge columns. In the two lowest windows of $(\Aosc, \omega)$ the grid ends at a smaller amplitude (Figs.~\ref{fig:profile2d} and \ref{fig:profile2d_wc}), since at larger amplitudes the optima of some nodes lie on the boundary of the region in which the emulator is used, far from the best fit. In the $(\Aosc, \phi)$ plane $\omega$ is free over its whole prior. However, the scan of this plane missed the narrow maxima at high frequency, where it was up to 23 lower in $\chi^2$ than the other two planes. We have therefore added studies of this plane with $\omega$ restricted to a narrow range around each of the highest maxima, four for \PDP{} and five for \HWC{}, and at each node the largest of the values is used.

The profile of the emulator itself, shown in Figs.~\ref{fig:omega} and \ref{fig:omega_wc}, agrees closely with the true profile. We compare the two at the nodes at which the refinement with the true likelihood met its convergence test, since at the other nodes the value calculated with the true likelihood is only a lower bound on the profile likelihood. Within $\Delta\chi^2 < 2.30$ of the best fit, the difference between the value of a node calculated with the true likelihood and the maximum of the emulator at that node has an rms of 0.04 in $\chi^2$ for \PDP{} and 0.10 for \HWC{}, and within $\Delta\chi^2 < 11.83$ the rms is 0.21 and 0.24.

The refinement with the true likelihood is therefore cheap where the emulator is accurate. At 90\% of the 20\,510 nodes of \PDP{} with $\Delta\chi^2 < 11.83$, the refinement lowers $\chi^2$ by less than 0.04 below its value at the optimum of the emulator, and it needs on average 27 evaluations of the true likelihood per node. 96\% of these nodes met the convergence test of the refinement, and 36 reached the maximum of 150 evaluations, all at $\Delta\chi^2 > 8.8$. Since the value of a node is the best value of the true likelihood found, a node that did not converge can only underestimate the profile likelihood. In total the studies of \PDP{} required \ensuremath{1.4\times10^{6}} evaluations of the true likelihood, and a study ran on a single cluster node with 46 processes. The two planes that contain $\omega$ give the same one-dimensional profile to within 0.20 in $\chi^2$. For \HWC{} there are 26\,869 such nodes, at 90\% of which the refinement lowers $\chi^2$ by less than 0.40, with on average 74 evaluations per node. 53\% of them met the convergence test, and 3\,712 reached the maximum of evaluations, all at $\Delta\chi^2 > 1.0$.

The best fits without oscillation are very well determined, since the nodes of the row $\Aosc = 0$ agree to within 0.07 in $\chi^2$ for \PDP{} and 0.11 for \HWC{}. Apart from the maxima discussed in Sec.~\ref{sec:results:profiles}, the profile of \PDP{} has maxima at $\omega \approx 100$ and 30, which are worse than the best fit by 3.6 and 4.3 in $\chi^2$, and at $\omega < 2.6$ the improvement reaches 3.0. In its $(\omega, \phi)$ plane the contour $\Delta\chi^2 = 2.30$ encloses the peak of the best fit and that at $\omega \approx 196$, and the contour $\Delta\chi^2 = 6.18$ encloses islands in 10 separate ranges of $\omega$. Note that these levels are the Gaussian ones for two parameters and serve only as a reference, since $\omega$ and $\phi$ are not defined at $\Aosc = 0$. For \HWC{} the maxima next to the best fit, at $\omega \approx 36$ and 22, are worse than it by 0.9 and 1.9 in $\chi^2$. The heights of the maxima of the profile in $\omega$ quoted in this paper are, as for the best fits, those of nodes of the $(\Aosc, \phi)$ plane, in which $\omega$ is not restricted to the grid. At the frequency and phase of the best fit of \PDP{} the data of \HWC{} prefers a small amplitude, $\Aosc = 0.005$, which improves $\chi^2$ by only 0.6.

The look-elsewhere estimate of Sec.~\ref{sec:results:comparison} counts the upcrossings of the one-dimensional profile of $\omega$ on the grid of the $(\omega, \phi)$ plane, whose step is about 1 in $\omega$. Counted at levels of the improvement between 0.5 and 3 instead of 1, the expected number of maxima as high as the best fit lies between 0.7 and 1.0 for \PDP{}, and between 2.4 and 3.4 for \HWC{}.

The one-dimensional profile of the amplitude gives $0.016 < \Aosc < 0.031$ for $\Delta\chi^2 < 1$ and $0.009 < \Aosc < 0.075$ for $\Delta\chi^2 < 4$ with \PDP{}, and $0.008 < \Aosc < 0.017$ and $0.003 < \Aosc < 0.084$ with \HWC{}. At the upper end of the grid, $\Aosc = 0.1$, the profile of \PDP{} is worse than the best fit by 10.6 in $\chi^2$. At $\omega = 1$, next to the lower prior boundary, the oscillation is compensated by $n_\mathrm{s}$ and the running (Sec.~\ref{sec:theory:template}), and there the profile comes within 8.0 of the best fit in $\chi^2$ for \PDP{} and 2.7 for \HWC{}, at amplitudes of about 0.04 to 0.05.

\subsection{The split of the prior volume}
\label{app:split}

\begin{table}[tb]
\centering
\small
\renewcommand{\arraystretch}{1.1}
\begin{tabular}{rcccl}
\cmidrule[\heavyrulewidth](lr){2-5}
 & Part & Training box in $\omega$ & Band to the next & Largest maxima of the scan \\
\cmidrule(lr){2-5}
\PDP{} & low & $[1, 3.1]$ & $2.6$--$3$ & none above 3 \\
 & 1a & $[2.5, 39.25]$ & $36.75$--$38.25$ & $21.5$ ($4.8$), $18.25$ ($4.6$), $30.5$ ($4.5$) \\
 & 1b & $[35.75, 63]$ & $60.5$--$62$ & $51.5$ ($10.2$) \\
 & 2a & $[59.5, 89.75]$ & $87.25$--$88.75$ & $64.75$ ($3.9$), $67.5$ ($3.3$) \\
 & 2b & $[86.25, 113]$ & $110.5$--$112$ & $100.75$ ($6.5$), $104.75$ ($5.8$), $97.75$ ($3.2$) \\
 & 3 & $[109.5, 156]$ & $153.5$--$155$ & $128.25$ ($4.2$), $137.25$ ($4.1$), $150.5$ ($3.4$) \\
 & 4 & $[152.5, 200]$ & -- & $196$ ($8.6$), $179$ ($8.0$), $182.25$ ($4.0$) \\
\cmidrule(lr){2-5}
\HWC{} & low & $[1, 6.1]$ & $6$--$6.1$ & not scanned \\
 & 1a & $[5.5, 39.25]$ & $36.75$--$38.25$ & not scanned \\
 & 1b & $[35.75, 63]$ & $60.5$--$62$ & $55.5$ ($4.1$), $57.5$ ($3.9$) (scanned from $55$) \\
 & 2a & $[59.5, 89.75]$ & $87.25$--$88.75$ & $81.5$ ($5.0$), $73$ ($3.7$) \\
 & 2b & $[86.25, 113]$ & $110.5$--$112$ & none above 3 \\
 & 3 & $[109.5, 156]$ & $153.5$--$155$ & $141.5$ ($5.6$), $146.25$ ($4.8$), $140$ ($3.9$) \\
 & 4a & $[152.5, 176.25]$ & $173.75$--$175.25$ & $157.75$ ($6.2$), $172$ ($4.9$), $168$ ($4.9$) \\
 & 4b & $[172.75, 200]$ & -- & $195.5$ ($5.0$), $193.5$ ($4.3$), $197.5$ ($3.4$) \\
\cmidrule[\heavyrulewidth](lr){2-5}
\end{tabular}
\caption{The parts into which the prior volume is split in $\omega$, with the training box of each emulator, the band in which it is blended with the next one, and the largest maxima of the scan of the likelihood in the range in which the emulator is used, with their improvements in $\chi^2$ at the reference point (Sec.~\ref{sec:theory:template}). The scan of \HWC{} starts at $\omega = 55$. All emulators of a dataset use the same likelihood and the same prior, $1 \leq \omega \leq 200$.}
\label{tab:zones}
\renewcommand{\arraystretch}{1.0}
\end{table}

The parts into which the prior volume is split, with the training boxes of their emulators, are listed in Table~\ref{tab:zones}. The split of \HWC{} differs from that of \PDP{} in two places. The lowest part ends at $\omega \approx 6$ instead of 3, and the zone at the highest frequencies is split once more, at $\omega = 174.5$, since the scan of \HWC{} has nine maxima of comparable height above $\omega = 155$. The emulators of a dataset are combined with the tool \texttt{combine} of \client{}, which checks that they emulate the same likelihood with the same parameters and that their training boxes differ only in $\omega$. The combined emulator is a single network. The parts are restricted to their own regions (Appendix~\ref{app:boosts}), and neighbouring parts are blended in the bands of Table~\ref{tab:zones}, which lie inside the training boxes of both. In a band the upper emulator has the weight $s = 6t^5 - 15t^4 + 10t^3$, where $t$ runs from 0 to 1 across the band, so that the first and second derivatives of the weight vanish at both ends, and outside the bands the value of the combined emulator is exactly that of one emulator. The region of the combined emulator is the union of the regions of the parts, and for the derived parameters we use the network of each part in the range in which its emulator is used. Inside the bands the improvement in $\chi^2$ of the scan of Sec.~\ref{sec:theory:template} is at most 1.1 for \PDP{} and 1.0 for \HWC{} above $\omega = 55$, compared with up to 10.2 at the maxima.

In all bands but the lowest, the combined emulator reproduces the true likelihood at the training points in the band, within 30 of the best point in $\ln\mathcal{P}$, with an rms error of at most 0.07 for \PDP{} and 0.14 for \HWC{}. In the lowest band of \PDP{} the emulator above the band is accurate only for small amplitudes. However, the band holds only 0.03\% of the posterior. The effect on the posterior and the evidence is therefore negligible, and the profile likelihood is affected only at $\Delta\chi^2 > 13$, outside the contours of Fig.~\ref{fig:profile2d}. For \HWC{} the band is placed at the top of the overlap, $6.0 < \omega < 6.1$, and there the combined emulator reproduces the true likelihood with an rms error of 0.14.

\section{Precision of \CLASSpp{} for an oscillating primordial spectrum}
\label{app:classpp}

The default precision settings of \CLASSpp{} are chosen for smooth primordial spectra, and at high frequency they do not resolve an oscillation in $\ln k$. Three samplings matter here: the table of the primordial spectrum in $k$, the grid of wavenumbers of the transfer functions (\texttt{q\_linstep\_trans} and \texttt{q\_logstep\_trans}), and the multipoles at which the spectra are computed and between which they are interpolated (\texttt{l\_logstep} and \texttt{l\_linstep}). The oscillation appears in the spectra with a period $\Delta\ell \approx 2\pi\ell/\omega$, which the sampling in $\ell$ must resolve.

We measure the precision on the likelihood itself. At the maxima of the scan of Sec.~\ref{sec:theory:template} we compare the improvement in $\chi^2$ due to the oscillation, $\chi^2(\Aosc = 0) - \chi^2(\Aosc, \omega, \phi)$, with that computed with a much denser sampling, the reference (logarithmic and linear steps of 1.01 and 3 in $\ell$, \texttt{q\_linstep\_trans}~$= 0.05$ and \texttt{q\_logstep\_trans}~$= 45$; Table~\ref{tab:classpp}). With the default sampling of the multipoles the improvement at $\omega = 50$ is only 0.7, compared with 8.5 with the reference. A sampling that is denser only in $\ell$ agrees with the reference to within 0.07 up to $\omega = 90$, but at $\omega = 100$ the improvement is already 0.3 too small, and a still denser sampling in $\ell$ does not change this. A comparison of the lensed spectra, changing one sampling at a time (Table~\ref{tab:classpp_cl}), shows that at higher frequencies both $\ell$ and $k$ must be sampled densely, that the primordial table then needs 1000 points per decade, and that the fast computation of the lensed spectra changes the imprint by up to 0.024 at $\omega \geq 160$ for \PDP{}.

\begin{table}[tb]
\centering
\small
\begin{tabular}{rcccccc}
\cmidrule[\heavyrulewidth](lr){2-7}
 & $\omega$ & $\Aosc$ & \multicolumn{4}{c}{Improvement in $\chi^2$ with the sampling} \\
\cmidrule(lr){4-7}
 & & & default & $\omega \lesssim 90$ & $\omega \leq 200$ & reference \\
\cmidrule(lr){2-7}
\PDP{} & $18.25$ & $0.009$ & $4.14$ & $4.59$ & -- & $4.59$ \\
 & $21.5$ & $0.009$ & $4.31$ & $4.74$ & -- & $4.74$ \\
 & $30.5$ & $0.010$ & $1.52$ & $4.53$ & -- & $4.53$ \\
 & $50$ & $0.021$ & $0.65$ & $8.51$ & -- & $8.47$ \\
 & $51.5$ & $0.024$ & -- & $10.25$ & -- & $10.27$ \\
 & $65$ & $0.017$ & -- & $3.80$ & -- & $3.83$ \\
 & $67.5$ & $0.016$ & -- & $3.38$ & -- & $3.31$ \\
 & $100$ & $0.024$ & -- & $4.30$ & -- & $4.61$ \\
 & $100.75$ & $0.029$ & -- & $6.48$ & $6.53$ & $6.53$ \\
 & $104.75$ & $0.028$ & -- & $5.37$ & $5.83$ & $5.82$ \\
 & $179$ & $0.047$ & -- & $6.67$ & $7.96$ & $7.88$ \\
 & $196$ & $0.048$ & -- & $2.45$ & $8.62$ & $8.72$ \\
 & $200$ & $0.036$ & -- & $0.88$ & $4.22$ & $3.73$ \\
\cmidrule(lr){2-7}
\HWC{} & $81.5$ & $0.024$ & -- & $4.86$ & $5.01$ & $5.01$ \\
 & $141.5$ & $0.046$ & -- & $6.72$ & $5.56$ & $5.62$ \\
 & $157.75$ & $0.054$ & -- & $-0.44$ & $6.20$ & $6.20$ \\
 & $195.5$ & $0.058$ & -- & $-5.07$ & $4.99$ & $4.82$ \\
 & $200$ & $0.024$ & -- & $0.01$ & $0.79$ & $0.54$ \\
\cmidrule[\heavyrulewidth](lr){2-7}
\end{tabular}
\caption{The improvement in $\chi^2$ due to the oscillation at the maxima of the scans of Sec.~\ref{sec:theory:template}, for four samplings of \CLASSpp{}: its default sampling of the multipoles, a sampling denser only in $\ell$ that suffices up to $\omega \approx 90$ ($\omega \lesssim 90$: logarithmic and linear steps of 1.026 and 8, for \HWC{} a linear step of 20 above $\ell \approx 770$), the settings of our runs ($\omega \leq 200$) and the reference. $\Aosc$ is the amplitude at the maximum. For \HWC{}, and for \PDP{} above $\omega = 100$, the reference also has a primordial table of 1000 points per decade and the accurate computation of the lensed spectra. The maximum at $\omega = 50$ is that of a first scan which ended at this frequency.}
\label{tab:classpp}
\end{table}

\begin{table}[tb]
\centering
\small
\begin{tabular}{rccccccc}
\cmidrule[\heavyrulewidth](lr){2-8}
 & $\omega$ & Imprint & \multicolumn{5}{c}{Error with the sampling} \\
\cmidrule(lr){4-8}
 & & & $\omega \lesssim 90$ & table & $\ell$ & $k$ & $\omega \leq 200$ \\
\cmidrule(lr){2-8}
\PDP{} & $60$ & $5.4$ & $0.009$ & $0.009$ & $0.009$ & $0.000$ & $0.000$ \\
 & $80$ & $3.8$ & $0.015$ & $0.015$ & $0.015$ & $0.001$ & $0.000$ \\
 & $100$ & $2.9$ & $0.033$ & $0.033$ & $0.027$ & $0.007$ & $0.000$ \\
 & $130$ & $2.1$ & $0.092$ & $0.091$ & $0.041$ & $0.071$ & $0.000$ \\
 & $160$ & $1.7$ & $0.252$ & $0.253$ & $0.049$ & $0.212$ & $0.001$ \\
 & $200$ & $1.5$ & $0.369$ & $0.380$ & $0.098$ & $0.293$ & $0.010$ \\
 & \multicolumn{2}{l}{time per spectrum [s]} & $1.6$ & $3.7$ & $7.2$ & $25.1$ & $11.1$ \\
\cmidrule(lr){2-8}
\HWC{} & $60$ & $10.1$ & $0.001$ & $0.001$ & $0.000$ & $0.001$ & $0.000$ \\
 & $100$ & $4.7$ & $0.088$ & $0.086$ & $0.013$ & $0.074$ & $0.000$ \\
 & $150$ & $2.8$ & $1.077$ & $1.086$ & $0.026$ & $1.064$ & $0.000$ \\
 & $200$ & $2.0$ & $1.971$ & $2.109$ & $0.092$ & $2.025$ & $0.006$ \\
 & \multicolumn{2}{l}{time per spectrum [s]} & $8.0$ & $8.4$ & $14.6$ & $57.2$ & $41.5$ \\
\cmidrule[\heavyrulewidth](lr){2-8}
\end{tabular}
\caption{The imprint of an oscillation with $\Aosc = 0.02$ on the lensed CMB spectra computed by \CLASSpp{}, and its error with five samplings, both as a noise-weighted $\chi^2$, with the noise of Planck for \PDP{} (TT up to $\ell = 2500$, TE and EE up to 2000, $f_\mathrm{sky} = 0.6$) and of ACT for \HWC{} ($30 \leq \ell \leq 6327$, $f_\mathrm{sky} = 0.3$). The samplings are $\omega \lesssim 90$ of Table~\ref{tab:classpp}, the same with a primordial table of 1000 points per decade (table), this with a denser sampling in $\ell$ (steps of 1.013 and 4 for \PDP{}, 8 for \HWC{}) or in $k$ (0.05 and 45), and the settings of our runs ($\omega \leq 200$, for \PDP{} here with the fast computation of the lensed spectra). The errors are relative to the reference of Table~\ref{tab:classpp} for \HWC{}, and to a sampling about twice as dense in $\ell$, $k$ and the table for \PDP{}. The last row of each dataset is the time of one computation on one core.}
\label{tab:classpp_cl}
\end{table}

In our runs we therefore use logarithmic and linear steps of 1.013 and 8 in $\ell$, \texttt{q\_linstep\_trans}~$= 0.15$ and \texttt{q\_logstep\_trans}~$= 90$ (the defaults are 0.45 and 400), a primordial table of 1000 points per decade and the accurate computation of the lensed spectra. The scan of Sec.~\ref{sec:theory:template} was computed with these settings above $\omega = 55$, and below with the sampling that suffices up to $\omega \approx 90$. With our settings the improvement in $\chi^2$ agrees with the reference to within 0.09 for \PDP{} and 0.18 for \HWC{} at all maxima below $\omega = 200$, i.e.\ to about a tenth in $\ln\mathcal{P}$, which is small compared with the error of the emulators (Appendix~\ref{app:accuracy}), and to within 0.5 at $\omega = 200$. With the sampling that suffices up to $\omega \approx 90$, on the other hand, the improvements at high frequency are wrong by up to 10, and some even become negative. An evaluation of the likelihood with our settings takes about 24 core seconds for \PDP{} and 78 for \HWC{}, roughly three to four times as long as with that sampling.

The likelihoods of ACT and SPT compare the theory with binned bandpowers. Comparing the $\chi^2$ of the imprint in the bandpowers with that in the same data multipole by multipole, we find that the binning keeps 57\% of it at $\omega = 60$ and 7\% at $\omega = 200$ for ACT, and 73\% and 3\% for SPT, while CamSpec, which compares the spectra multipole by multipole, loses nothing. With the Planck spectra used by the two datasets, the $\chi^2$ of the imprint of $\Aosc = 0.02$ at $\omega = 200$ is 1.5 for \PDP{} and 0.7 for \HWC{}, 81\% of which comes from the Planck spectra of \HWC{}.

\section{Computational cost compared with standard methods}
\label{app:cost}

The analysis of this paper could not have been carried out by calling \CLASSpp{} directly, mainly because of the wall time. An evaluation of the likelihood takes 24 core seconds for \PDP{} and 78 for \HWC{}. In \client{} the evaluations are independent, and the points of a boost are computed on all 96 cores of a run at the same time. In a Markov chain, on the other hand, a new point depends on the previous one, more chains help only up to about 6--8, since each must converge by itself, and \CLASSpp{} does not run efficiently on more than a few cores. The wall time of a standard analysis is therefore set by the number of evaluations that must be made in sequence, and not by the size of the cluster.

To quantify this we have run standard samplers on the combined emulator of \HWC{} in place of \CLASSpp{} and counted the evaluations of the likelihood (Table~\ref{tab:cost}). Since the emulator reproduces the true likelihood closely (Appendix~\ref{app:accuracy}), a sampler should need roughly the same number of evaluations with \CLASSpp{}, although the numerical noise of \CLASSpp{} is not included. With the Metropolis--Hastings sampler of \cobaya{} \cite{Lewis:2002ah,Lewis:2013hha,Torrado:2020dgo}, started from the covariance of the posterior of the emulator, a posterior converged to $R-1 = 0.01$ \cite{Gelman:1992zz} over the whole prior would take about 3 months. \client{} needed 23 h for a posterior and evidence that had stopped changing after five boosts (Appendix~\ref{app:boosts}).

The evidence requires many more evaluations. With nested sampling as in PolyChord \cite{Handley:2015fda,Handley:2015vkr}, run in dynesty \cite{Speagle:2019ivv}, a new live point takes about 600 evaluations. Since $\sigma(\ln Z) \simeq \sqrt{H/n_\mathrm{live}}$, with $H = 37$ nats, an error of one requires only 37 live points and \ensuremath{\sim 1.2\times10^{6}} evaluations. However, such a run would very likely miss some of the many maxima in $\omega$, and it would not determine the Bayes factor of Sec.~\ref{sec:actspt:profiles}, $\ln B = -1.2$. With the 2048 live points of our evidence runs the error is 0.13, and nested sampling would need \ensuremath{\sim 6.4\times10^{7}} evaluations, or \ensuremath{\sim 1.4\times10^{6}} core hours, which would take about 3 months on the 768 cores of our runs. This is roughly 110 times the evaluations that \client{} needed for the posterior and the evidence.

For the profile likelihoods we compare with PROSPECT \cite{Holm:2023uwa}, which maximises by simulated annealing without gradients. For four nodes of the profile of \PDP{} with $38 \leq \omega \leq 60$, started from Metropolis--Hastings chains restricted to this range as in ref.~\cite{Holm:2023uwa}, it came within 0.06 in $\chi^2$ of the profile of the emulator with 1\,040 to 40\,040 evaluations per node. Without such chains it must be started near each of the about 50 maxima of the profile in $\omega$ (Fig.~\ref{fig:omega}): started from five values of $\omega$ in this range, the annealing ended at the maximum nearest to its start, and in four of the cases 9.0 and 11.5 below the profile in $\chi^2$. \client{} refines a node of \HWC{} with on average 86 evaluations of the true likelihood over all nodes of the surfaces, since the refinement starts at the optimum of the emulator. The 2500 nodes of a surface therefore cost about as many evaluations as 5 to 210 nodes with PROSPECT, without the chains from which these start. The mean of 74 evaluations in Appendix~\ref{app:profiles} is that of the nodes within $\Delta\chi^2 < 11.83$ of the best fit.

In conclusion, with standard methods the posterior and the profile likelihoods of \HWC{} would take months of wall time, and an evidence as precise as ours \ensuremath{\sim 1.4\times10^{6}} core hours. With \client{} a robust posterior and evidence took 23 h and \ensuremath{1.6\times10^{4}} core hours, and the profile likelihoods took days.

\begin{table}[tb]
\centering
\small
\begin{tabular}{>{\raggedright\arraybackslash}p{3.2cm}>{\raggedright\arraybackslash}p{3.4cm}rrr}
\cmidrule[\heavyrulewidth](lr){1-5}
\textbf{Goal} & \textbf{Method} & \textbf{Evaluations} & \textbf{Core hours} & \textbf{Wall time} \\
\cmidrule(lr){1-5}
Posterior & Metropolis--Hastings & $\sim 3\times10^{6}$ & $\sim 6\times10^{4}$ & 3 months \\
Evidence, $\sigma(\ln Z) \approx 1$ & Nested sampling, 37 live points & $\sim 1.2\times10^{6}$ & $\sim 2.5\times10^{4}$ & 5 days \\
Evidence, $\sigma(\ln Z) = 0.13$ & Nested sampling, 2048 live points & $\sim 6.4\times10^{7}$ & $\sim 1.4\times10^{6}$ & 3 months \\
Profile likelihood, one node & Simulated annealing (PROSPECT) after Metropolis--Hastings & $1\,040$--$4.0\times10^{4}$ & 23--869 & 4 h--6 days \\
\cmidrule(lr){1-5}
Posterior and evidence & \client{}, five boosts & $6.0\times10^{5}$ & $1.6\times10^{4}$ & 23 h \\
Posterior and evidence & \client{}, ten boosts (as run) & $1.1\times10^{6}$ & $3.1\times10^{4}$ & 2 days \\
Profile likelihood, one node & \client{}, refinement (mean) & 86 & 1.9 & \\
2D profile likelihood, 2500 nodes & \client{}, refinement & $2.2\times10^{5}$ & $4\,676$ & \\
\cmidrule[\heavyrulewidth](lr){1-5}
\end{tabular}
\caption{The evaluations of the likelihood, the core hours and the wall time required for the posterior, the evidence and the profile likelihoods of \HWC{} with standard methods calling \CLASSpp{} directly, compared with \client{}. The Metropolis--Hastings posterior is taken to require at least the evaluations of chains restricted to $38 \leq \omega \leq 60$, and a value with $\sim$ is extrapolated, for the chains with $R-1 \propto 1/N$. Nested sampling is assumed to stop when the live points hold 0.1\% of the evidence; stopping when they hold half of it would save 24\% of the evaluations. The simulated annealing was run for \PDP{}. The core hours are the evaluations times 78 core seconds for the standard methods and for the refinement of the profiles, and for the posterior and the evidence of \client{} those held by its runs, training included. The wall times assume each evaluation on 4 cores with a speed-up of 3, 8 Metropolis--Hastings chains, nested sampling on 768 cores with the parallel speed-up of PolyChord \cite{Handley:2015vkr}, and one annealing chain per node of a profile. Those of \client{} are the times of the runs, with 96 cores for each emulator, all trained at the same time.}
\label{tab:cost}
\end{table}

\FloatBarrier

\bibliographystyle{JHEP}
\bibliography{Bibliography/references}

\end{document}